\documentclass[a4paper, prd,aps,tightenlines,superscriptaddress,showpacs,nofootinbib,eqsecnum, preprint]{revtex4}

\usepackage[british]{babel}
\usepackage{float}
\floatstyle{boxed}

\usepackage{graphicx,bm,dcolumn,amsmath,amsfonts,subfigure,color,latexsym,amssymb}

\def\@rcsid{\relax}
\def\rcsid#1{\def\next##1#1{\def\@rcsid{\mbox{RCS ##1}}}\next}
\def\M{{\cal M}}
\def\e{\eta}

\def\ba{\begin{eqnarray}}
\def\ea{\end{eqnarray}}
\def\be{\begin{equation}}
\def\ee{\end{equation}}

\def\nn{\nonumber}

\begin{document}


\draft

\title{Phenomenology of amplitude-corrected post-Newtonian gravitational waveforms for compact binary inspiral.\\ 
I. Signal-to-noise ratios}
\author{Chris Van Den Broeck}
\email{Chris.van-den-Broeck@astro.cf.ac.uk}
\author{Anand S.~Sengupta}
\email{Anand.Sengupta@astro.cf.ac.uk}

\affiliation{School of Physics and Astronomy, Cardiff University,\\
Queen's Buildings, The Parade, Cardiff CF24 3AA, United Kingdom}

\begin{abstract}

We study the phenomenological consequences of amplitude-corrected post-Newtonian (PN) gravitational waveforms, as opposed to the more commonly used restricted PN waveforms, for the quasi-circular, adiabatic inspiral of compact binary objects.  In the case of initial detectors it has been shown that the use of amplitude-corrected waveforms for detection templates would lead to significantly lower signal-to-noise ratios (SNRs) than those suggested by simulations based exclusively on restricted waveforms. We further elucidate the origin of the effect by an in-depth analytic treatment. The discussion is extended to advanced detectors, where new features emerge. Non-restricted waveforms are linear combinations of harmonics in the orbital phase, and in the frequency domain the $k$th harmonic is cut off at $k f_{LSO}$, with $f_{LSO}$ the orbital frequency at the last stable orbit. As a result, with non-restricted templates it is possible to achieve sizeable signal-to-noise ratios in cases where the dominant harmonic (which is the one at twice the orbital phase) does not enter the detector's bandwidth. This will have important repercussions on the detection of binary inspirals involving intermediate-mass black holes. For sources at a distance of 100 Mpc, taking into account the higher harmonics will double the mass reach of Advanced LIGO, and that of EGO  gets tripled. Conservative estimates indicate that the restricted waveforms underestimate detection rates for intermediate mass binary inspirals by at least a factor of twenty.

\end{abstract}

\pacs{04.25.Nx, 04.30.-w, 04.80.Nn, 95.55.Ym}

\maketitle

\section{Introduction}

A number of interferometric gravitational-wave detectors (LIGO, VIRGO, GEO600 and TAMA) are now running \cite{Observatories}, and the LIGO detector has reached design sensitivity. In the coming decade the Initial LIGO detector will be upgraded to Advanced LIGO. For all such detectors, the inspiral signals from neutron star and/or black hole binaries are among the strongest expected sources (see e.g.~\cite{Grishchuketal} for a review). In the late stages of inspiral, gravitational radiation backreaction will have circularized the components' orbits, and before the final plunge there will be an ``adiabatic" regime where the period of a single orbit is much shorter than the inspiral timescale. This part of the inspiral is relatively ``clean" and well-understood, so that the waveforms it produces can be modeled with great precision. This has been done in the so-called post-Newtonian (PN) approximation to general relativity, where waveforms are expressed as expansions in the orbital velocity $v$ (see \cite{Blanchet} for a review and extensive references).

The best waveforms currently available are of order $v^5$ in amplitude \cite{2.5PN} and $v^7$ in phase \cite{3.5PN}, which in the usual notation corresponds to 2.5PN and 3.5PN orders, respectively. These take the form of linear combinations of harmonics in the orbital phase. The second harmonic is the one that will usually dominate; it is the only one with a zeroth-order PN contribution to its amplitude. In most of the literature on adiabatic inspiral, one uses the \emph{restricted} post-Newtonian approximation \cite{last3minutes}, where all amplitude corrections are discarded and only PN contributions to the phase are taken into account. Thus, the restricted PN waveform consists of just the second harmonic with a prefactor at zeroth PN order.

An efficient way of searching for inspiral signals in data is matched filtering \cite{Helstrom}, which involves a bank of templates. Almost all templates currently used are based on the restricted PN approximation. The same kinds of waveforms that go into template banks are also used as simulated signals injected into stretches of data to evaluate algorithms that search for real events and veto spurious ones. These include the standard restricted PN waveforms, the so-called P-approximants \cite{comparison}, and the effective one-body (EOB) waveforms \cite{EOB}. The latter two result from resummation schemes designed to improve on the convergence of the phasing. The phenomenological templates for binary black hole inspiral that were recently proposed by Buonanno, Chen, Pan, and Vallisneri \cite{BCV} do have some simple corrections to their amplitudes, but they are still quite limited compared to the full amplitude-corrected waveforms. Overall the emphasis has been on the evolution of the phase rather than the amplitude, due to the belief that it is more important to know the phasing and number of cycles (or more precisely the number of useful cycles \cite{UsefulCycles}) of the signal in the detector's bandwidth. 

To our knowledge it has never been investigated in detail to what extent the use of restricted waveforms as templates and simulated signals is justified beyond heuristic considerations. Sintes and Vecchio \cite{SV} did study the effect of 0.5PN amplitude corrections on signal-to-noise ratios (SNRs). However, as shown in \cite{letter}, this happens to be a very special case; in going from 0PN to 0.5PN in amplitude, SNRs will increase, but from 1PN onwards there will be a pronounced drop, at least for stellar mass inspirals as seen in initial detectors. The effect of 2PN amplitude corrections on parameter estimation was studied by Hellings and Moore \cite{HellingsMoore}, but specifically in the context of LISA and without discussing the effect of higher harmonics on detection rates for systems with various masses. Here we investigate the use of non-restricted versus restricted waveforms for the purposes of detection, focusing on three ground-based detectors: Initial LIGO, Advanced LIGO, and a possible third-generation detector called the European Gravitational-Wave Observatory (EGO) \cite{Punturo}.  

Given a waveform $h$, the signal-to-noise ratio (SNR) for the ``detection" of $h$ using its normalized counterpart as a template is given by $\rho[h] \equiv (h|h)^{1/2}$, where $(\,.\,|\,.\,)$ is the usual inner product in terms of the noise power spectral density $S_h(f)$ of the detector. With the convention $\tilde{x}(f) = \int_{-\infty}^{\infty} x(t)\,\exp(-2\pi i f t)\,dt$ for Fourier transforms,
\be
(x|y)  \equiv 4\int_{f_s}^{f_{end}} \frac{\mbox{Re}[\tilde{x}^\ast (f)\tilde{y}(f)]}{S_h(f)}df.
\label{innerproduct}
\ee

Let $h_0$ be a restricted waveform and $h$ a waveform that is of 2.5PN order in amplitude and 3.5PN in phase, with both having the same parameters. With the above notation, $\rho_[h_0]$ is the kind of SNR one encounters in current simulated searches. In the future one may want to consider amplitude-corrected waveforms $h$ both for templates and simulated signals, in which case one would arrive at SNRs $\rho[h]$. As indicated in an earlier paper \cite{letter} in the context of initial detectors and stellar mass binaries, one has 
\be
\rho[h] < \rho[h_0]. 
\label{overestimation}
\ee
Hence, surprisingly, if one were to use the best available amplitude-corrected waveforms for detection templates, one should expect SNRs in actual searches to be lower than those suggested by simulations based purely on restricted PN waveforms. In Initial LIGO this overestimation can be as large as 25\% depending on what the parameter values are. Because SNR is inversely proportional to distance, this corresponds to an overestimation of the accessible spatial volume, and hence the detection rate, by up to a factor of two. Moreover, SNRs exhibit a downward trend as the post-Newtonian order of the amplitudes is increased from 1PN to 2.5PN in steps of 0.5PN. Should this trend continue then the overestimation would be worse still when taking into account amplitude corrections beyond the highest PN order currently available.
A rough indication of the origin of (\ref{overestimation}) was already given in \cite{letter}. The effect can be traced to the PN amplitude corrections of the dominant harmonic. Here we will provide a much more detailed analysis.

The situation is different for second and third generation detectors. In the frequency domain, the third and higher harmonics in amplitude-corrected waveforms are cut off at higher frequencies than the dominant one. Because the cut-off frequencies are inversely proportional to total mass, this means that higher harmonics may still enter the detector's bandwidth even if the dominant harmonic does not. In advanced detectors these harmonics can lead to a sizeable SNR. If the second harmonic does not enter the bandwidth, or if it enters only in a small frequency interval, the inequality (\ref{overestimation}) is reversed. This opens up the possibility of seeing intermediate-mass inspiral events with higher total mass than one would expect on the basis of the restricted waveforms. Indeed, the inclusion of amplitude-corrected higher harmonics in detection templates would effectively double the accessible mass range of Advanced LIGO, and it would triple that of EGO. For intermediate-mass inspirals the detection rates from non-restricted waveforms are several orders of magnitude larger than the ones computed from the restricted waveforms. Stellar mass inspirals would be detectable with EGO throughout a significant part of the visible Universe.

This paper is structured as follows. In section \ref{s:waveforms} we discuss the amplitude-corrected waveforms, as seen in a detector, up to 2.5PN in amplitude and 3.5PN in phase in the stationary phase approximation. In section \ref{s:SNRs} we compare signal-to-noise ratios using restricted and non-restricted waveforms as templates and simulated signals in Initial LIGO. The numerical results are explained analytically up to 1PN in amplitude. We then focus on the advantages of non-restricted templates in advanced detectors (section \ref{s:advanced}), including a third generation detector called EGO. Finally we summarize our results and conclusions. Expressions for the harmonics in the amplitude-corrected waveform can be found in Appendix A. We will go into some detail, outlining the general form of the harmonics up to 2.5PN in amplitude and giving explicit expressions up to 1PN.\footnote{The general expression for the complete waveform in the stationary phase approximation up to 2.5PN in amplitude is available upon request as a Mathematica notebook.} These expressions are also to be the basis for future work \cite{paramest}, where we will use the amplitude-corrected waveforms for parameter estimation.

Sensitivity curves for detectors will be taken as in \cite{Arunetal}, except for the EGO noise curve which can be found in Appendix C. Unless stated otherwise we use units such that $G=c=1$. We use the notation $(p,q)$PN for waveforms that have PN corrections up to order $p$ in amplitude and $q$ in phase. Integrals of the form (\ref{innerproduct}) can not be performed analytically and were obtained numerically using the software package Mathematica. Finally, when plotting quantities against mass in a cosmological context, we will give precedence to physical mass over (redshifted) observed mass. This is because we will encounter relatively large redshifts, in which case the use of observed mass would obscure the astrophysical consequences of the results.

\section{Amplitude-corrected post-Newtonian waveforms}
\label{s:waveforms}

The waveforms in the two polarizations take the general form
\be
h_{+,\times}=\frac{2M\eta}{r} x \,
\left\{H^{(0)}_{+,\times} + x^{1/2}H^{(1/2)}_{+,\times} + x H^{(1)}_{+,\times}
+ x^{3/2}H^{(3/2)}_{+,\times} + x^2 H^{(2)}_{+,\times} + x^{5/2} H^{(5/2)}_{+,\times}
\right\} 
\label{hpluscross}
\ee
where $r$ is the distance to the binary, $M$ its total mass, and $\e$ the ratio of reduced mass to total mass. The post-Newtonian expansion parameter is defined as $x=(2\pi M F)^{2/3} = v^2$, with $F(t)$ the instantaneous orbital frequency. The coefficients $H^{(n/2)}_{+,\times}$, $n=0, \ldots, 5$, are linear combinations of various harmonics with prefactors that depend on the inclination angle $\iota$ of the angular momentum of the binary with respect to the line of sight as well as on $\e$; their explicit expressions can be found in \cite{2.5PN}. The measured signal also depends on the polarization angle and the position in the sky through the detector's beam pattern functions $F_{+,\times}$:
\be
h(t)=F_+ h_+(t) + F_\times h_\times(t). \label{signal}
\ee
Note that for ground-based detectors, which are the ones we will be concerned with, it is reasonable to approximate $F_{+,\times}$ as being constant over the duration of the signal. They depend on 
angles $(\theta,\phi,\psi)$, where $(\theta,\phi)$ determine sky position while $\psi$ is the polarization 
angle. The signal (\ref{signal}) is a linear combination of harmonics of the orbital phase $\Psi(t)$ with offsets $\varphi_{(k,m/2)}$:
\be
h(t) = \sum_{k=1}^{N_p} \sum_{m=0}^{2p} A_{(k,m/2)}(t) \cos(k\Psi(t) + \varphi_{(k,m/2)}),
\label{lincomb}
\ee
where the coefficients $A_{(k,m/2)}$ are functions of $(r,M,\e,\theta,\phi,\psi,\iota)$ multiplied by $x^{(m+2)/2}$. The orbital phase $\Psi(t)$ is a series in $x$, which in the case of non-spinning binaries is known to 3.5PN order. The number of harmonics $N_p$ depends on the PN order in amplitude, $p$; at 2.5PN one has $N_p=7$.

More explicitly, the waveform $h(t)$ is found as follows: 
\begin{itemize}
\item Substitute the expressions (\ref{hpluscross}) into (\ref{signal}), taking the prefactors $H^{(n/2)}_{+,\times}(\e,\iota)$ to be as in \cite{2.5PN}, and beam pattern functions
\ba
F_+(\theta,\phi,\psi) &=&  \frac{1}{2}\left(1+\cos^2(\theta)\right)\cos(2\phi)\cos(2\psi) 
- \cos(\theta)\sin(2\phi)\sin(2\psi), \nn\\
F_\times(\theta,\phi,\psi) &=& \frac{1}{2}\left(1+\cos^2(\theta)\right)\cos(2\phi)\sin(2\psi) 
+ \cos(\theta)\sin(2\phi)\cos(2\psi).
\label{beampatternfunctions}
\ea
\item In the resulting expression for $h(t)$, collect cosines and sines of multiples of the orbital phase to arrive at
\be
h(t) = \sum_{k=1}^{N_p} \sum_{m=0}^{2p} \left[\alpha_{(k,m/2)}(t) \cos(k\Psi(t)) + \beta_{(k,m/2)}(t) \sin(k\Psi(t)) \right], 
\ee
where the $\alpha_{(k,s)}$, $\beta_{(k,s)}$ depend on $(r,M,\e,\iota,\theta,\phi,\psi)$, as well as on time through $x$.
\item Combine corresponding sines and cosines into simple cosines to arrive at (\ref{lincomb}), 
with 
\be
A_{(k,s)} = \mbox{sign}(\alpha_{(k,s)})\,\sqrt{\alpha_{(k,s)}^2+\beta_{(k,s)}^2}
\ee
and
\be
\varphi_{(k,s)} = \tan^{-1}\left(-\frac{\beta_{(k,s)}}{\alpha_{(k,s)}}\right).
\ee
\end{itemize}

During the inspiral phase one has $|d\ln A_{(k,s)}/dt| \ll 1$ and 
$|k d^2\Psi/dt^2| \ll (k d\Psi/dt)^2$, in which case one can use the well-known stationary phase approximation (SPA) \cite{SPA} for the Fourier transform of (\ref{lincomb}):
\ba
\tilde{h}^{(k)}(f) &\simeq& \frac{\sum_{m=0}^5 A_{(k,m/2)}\left(t(\frac{1}{k}f)\right)}{2\sqrt{k\dot{F}\left(t\left(\frac{1}{k}f\right)\right)}}
                \exp\left[i\left(2\pi f t\left(\frac{1}{k}f\right) - k\Psi\left(t\left(\frac{1}{k}f\right)\right) - \varphi_{(k,m/2)} - \pi/4\right)\right] \nn\\
               &=& \frac{\sum_{m=0}^5 A_{(k,m/2)}\left(t(\frac{1}{k}f)\right)\,e^{-i\varphi_{(k,m/2)}}}{2 \sqrt{k\dot{F}\left(t(\frac{1}{k}f)\right)}} 
\exp\left[i\left(2\pi f t_c -\pi/4 + k \psi\left(\frac{1}{k} f\right)\right)\right]. \label{h_k} 
\ea
A dot denotes derivation w.r.t.~time and $t_c$ is the coalescence time. The function $F(t)$ is the instantaneous frequency associated with the orbital phase, $2\pi F(t) = \dot{\Psi}(t)$, and $t(f)$ is defined implicitly by
$F(t(f)) = f$. The expression for the ``frequency sweep" $\dot{F}$ in terms of $F$ is given in Appendix A. To 3.5PN order, the function $\psi(f)$ takes the form
\be
\psi(f) = -\psi_c + \frac{3}{256\,(2\pi \M f)^{5/3}}\sum_{i=0}^7 \psi_i (2\pi M f)^{i/3}
\label{phase}
\ee 
where $\psi_c$ is the orbital phase at coalescence, and for the coefficients $\psi_i$ we again refer to Appendix A.\footnote{Note that we are using slightly different conventions from e.g.~\cite{comparison,PRD66,Arunetal}, where the focus was on the dominant harmonic and $F(t)$ referred to \emph{twice} the orbital frequency. Here primacy is given to the orbital quantities themselves, so that the dependence of $\dot{F}$ on $F$ and the frequency dependence of $\psi$ look a little different from what one often finds in the literature.}

The SPA for the $(p,3.5)$PN
waveform is then
\be
\tilde{h}_{SPA}(f) = \sum_{k=1}^{N_p} \left[\frac{\sum_{m=0}^{2p} A_{(k,m/2)}\left(t\left(\frac{1}{k}f\right)\right)\,e^{-i\varphi_{(k,m/2)}}}{2\sqrt{k \dot{F}\left(t\left(\frac{1}{k} f\right)\right)}}\right]_p 
\exp\left[i\left(2\pi f t_c -\pi/4 + k \psi\left(\frac{1}{k} f\right)\right)\right], 
\label{SPA}
\ee
where $[\,.\,]_p$ denotes consistent truncation to $p$th post-Newtonian order (i.e., the ``Newtonian" prefactor $f^{-7/6}$ is taken outside and the remaining expression is expanded in $(2\pi M f)^{1/3}$ up to $(2\pi M f)^{2p/3}$).

The SPA is just one method of approximating the Fourier transform; the main alternative is the fast Fourier transform (FFT). However, as shown in \cite{SPAvsFFT,UsefulCycles}, the difference between the two in terms of overlap and SNRs tends to be only a few percent for restricted waveforms, at least in initial detectors, and the situation improves dramatically as the detector's sensitivity at low frequencies is increased. We also note that the conditions for the applicability of the stationary phase approximation on the harmonics become more favorable with increasing $k$. The reason why we prefer to work with the SPA is that it will give us direct analytic insight into the effect on detection rates of the presence or absence of amplitude corrections.

Finally, the expressions for the respective harmonics should not be used up to arbitrarily large frequencies; some maximum orbital frequency must be chosen at which waveforms are truncated. In the limit where one of the component masses is sent to zero while keeping total mass fixed, there is a well-defined last stable orbit with orbital frequency 
\be
f_{LSO} = \frac{1}{6^{3/2} 2\pi M}, \label{fLSO}
\ee
where $M$ is total mass. For binary systems with comparable component masses the situation is more involved \cite{LSO}, but for simplicity we adopt (\ref{fLSO}) as a formal expression also in that case. In the time domain we will consider the waveform to be valid only up to a time determined by $F(t) = f_{LSO}$. In the frequency domain this roughly corresponds to cutting off the $k$th harmonic at a frequency $f = k f_{LSO}$. To reflect this restriction on the validity of the harmonics, in practice we multiply the $k$th harmonic by $\theta(k f_{LSO}-f)$, where $\theta(x)$ is the usual Heaviside function ($\theta(x) = 0$ for $x < 0$ and 1 otherwise). In the definition of the inner product $(\,.\,|\,.\,)$, Eq.~(\ref{innerproduct}), we then have $f_{end} = 7 f_{LSO}$, the frequency reach of our highest harmonic. The fact that higher harmonics have a higher frequency reach will have important implications, especially for advanced detectors. The lower cut-off frequency $f_s$ depends on the detector. For Initial LIGO, Advanced LIGO, and EGO, we will take it to be 40 Hz, 20 Hz, and 10 Hz, respectively (see Appendix C).

\section{Signal-to-noise ratios: Initial detectors}
\label{s:SNRs}

We now turn to the issue of signal-to-noise ratio in Initial LIGO. Apart from binary neutron star inspirals, the inspiral events of main interest in the case of initial detectors are those involving stellar mass black holes. These result from massive main-sequence binary stellar systems (field binaries) or from capture processes in galactic centers or globular clusters (capture binaries). Field binaries are expected to have total masses in the range $10\,M_\odot \lesssim M \lesssim 50\,M_\odot$ while capture binaries may be somewhat heavier \cite{QuinlanShapiro}. Coalescence rate estimates for field binaries range from $\sim 10^{-8}\,\mbox{yr}^{-1}$ to $\sim 10^{-6}\,\mbox{yr}^{-1}$ in our galaxy, corresponding to $\sim 10^{-3}$--$10^{-1}\mbox{yr}^{-1}$ within a distance of 100 Mpc \cite{TutukovYungelson,Lipunovetal,Grishchuketal}. For capture binaries in globular clusters, coalescence rates have been estimated at $\sim 3\,\mbox{yr}^{-1}$ within 600 Mpc \cite{SigurdssonHernquist,FlanaganHughes}.

Below the implications of using non-restricted versus restricted waveforms for templates and simulated signals are first investigated numerically and then explained analytically.

\subsection{Numerical observations}

First let us look at the SNRs for inspirals at 20 Mpc in Initial LIGO. We consider three kinds of systems: a binary neutron star (NS--NS), a neutron star and a black hole (NS--BH), and a binary black hole (BH--BH). For concreteness we take the neutron star mass to be $M_{NS} = 1.4\,M_\odot$ while for the black hole $M_{BH}=10\,M_\odot$. Table \ref{t:PNdependence} displays the behavior of $\rho[h_0]$ and $\rho[h]$ for $(p,3.5)$PN waveforms $h$ with increasing $p$, with $h_0$ the corresponding restricted waveform. In going from $p=0$ to $p=0.5$, for the asymmetric system $\rho[h]$ increases slightly (as has been noted by Sintes and Vecchio \cite{SV}) while for the NS--NS and BH--BH there is no appreciable difference. However, at $p=1$, $\rho[h]$ drops below $\rho[h_0]$, and as $p$ is further increased a clear downward trend is seen.
At $p=2.5$ one has $\rho[h] < \rho[h_0]$, with the two differing by 3.6\%, 13.1\%, and 21.0\% for the NS--NS, NS--BH, and BH-BH inspirals, respectively. 

\begin{table}[htp!] 
\begin{tabular}{lrcccrcccrccc} 
\hline 
\hline 
&\vline& & NS--NS & &\vline& & NS--BH & &\vline& & BH--BH & \\
\hline
$p$ &\vline& \,\,\,$\rho[h_0]$\,\,\, && \,\,\,$\rho[h]$ &\vline& 
\,\,\,$\rho[h_0]$\,\,\,&& \,\,\,$\rho[h]$ &\vline& 
\,\,\,$\rho[h_0]$\,\,\, && \,\,\,$\rho[h]$\\
\hline 
0   &\vline& 6.465 && 6.465 &\vline& 13.492 && 13.492 &\vline& 30.928 && 30.928 \\
0.5 &\vline& "     && 6.465 &\vline&  "     && 13.936 &\vline& "      && 30.928 \\ 
1   &\vline& "     && 6.286 &\vline&  "     && 12.563 &\vline& "      && 28.140 \\
1.5 &\vline& "     && 6.286 &\vline&  "     && 12.417 &\vline& "      && 28.140 \\
2   &\vline& "     && 6.249 &\vline&  "     && 12.091 &\vline& "      && 26.377 \\
2.5 &\vline& "     && 6.238 &\vline&  "     && 11.933 &\vline& "      && 25.562 \\
\hline
\end{tabular}
\caption{Change in signal-to-noise ratios with increasing $p$ in $(p,3.5)$PN waveforms, for three different systems at 20 Mpc. (Angles were chosen arbitrarily as $\theta=\phi=\pi/6$, $\psi=\pi/4$, $\iota=\pi/3$.)} 
\label{t:PNdependence}
\end{table}

\begin{table}[htp!] 
\begin{tabular}{lrcrc} 
\hline 
\hline 
$(\theta,\phi)$ &\vline& $(\iota,\psi)$ &\vline& $\epsilon$ \\
\hline
$(0,0)$ &\vline& $(0,0)$ &\vline& 0.129 \\
\hline
$(\pi/6,\pi/3)$ &\vline& $(\pi/6,0)$ &\vline& 0.127 \\
"         &\vline& $(\pi/3,\pi/8)$ &\vline& 0.132 \\
"         &\vline& $(\pi/2,\pi/4)$ &\vline& 0.141 \\
\hline
$(\pi/3,\pi/3)$ &\vline& $(\pi/6,0)$ &\vline& 0.127 \\
"             &\vline& $(\pi/3,\pi/8)$ &\vline& 0.132 \\
"             &\vline& $(\pi/2,\pi/4)$ &\vline& 0.141 \\
\hline 
$(\pi/2,0)$    &\vline& $(\pi/6,0)$ &\vline& 0.127 \\
"             &\vline& $(\pi/3,\pi/8)$ &\vline& 0.128 \\
\hline
\end{tabular}
\caption{Angular dependence of the ratio $\epsilon$ for a NS--BH system as in Table \ref{t:PNdependence}.} 
\label{t:AngularDependence}
\end{table}

\begin{figure}[htbp!]
\centering
\includegraphics[scale=0.50,angle=0]{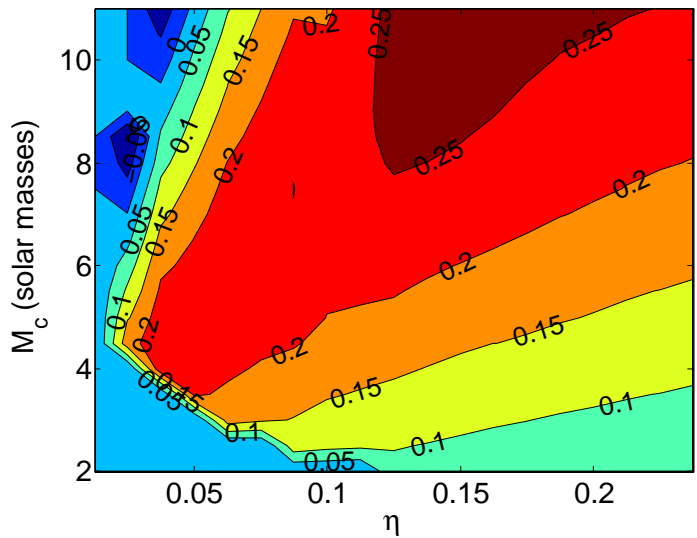}
\caption{A plot of $\epsilon$, the fractional overestimation of SNR due to the use of restricted PN waveforms, as a function of $\eta$ and chirp mass. (The sudden drop in the bottom left corner is due to the fact that we imposed $M >  1 \,M_\odot$.) In an important part of parameter space the overestimation is more than 20\%. However, also note the negative values for very massive, asymmetric systems.}
\label{f:SNR_Overestimation}
\end{figure}  

Thus, modeling templates and signals as restricted waveforms overestimates the SNRs. The fractional overestimation is given by
\be
\epsilon \equiv \frac{\rho[h_0]}{\rho[h]} - 1,
\ee
where from now on we let $h$ denote a $(2.5,3.5)$PN waveform.

In Table \ref{t:PNdependence} specific choices were made for the angles $(\theta,\phi,\iota,\psi)$ that encode the position and orientation of the source. Although $\rho[h_0]$ and $\rho[h]$ will vary significantly as functions of these angles, this is not the case for $\epsilon$. Indeed, as can be seen in Table \ref{t:AngularDependence}, for a NS--BH system the relative difference between $\rho[h_0]$ and $\rho[h]$ has only a very weak angular dependence. 

In Fig.~\ref{f:SNR_Overestimation} we have plotted $\epsilon$ as a function of $\eta$ and chirp mass $\mathcal{M} = M \eta^{3/5}$. In a significant part of parameter space the overestimation is above 20\%. Indeed, on the line corresponding to $\epsilon=0.2$, lighter and heavier component masses $m_1$, $m_2$ are constrained as $1\,M_\odot \lesssim m_1 \lesssim 12\,M_\odot$ and $12\,M_\odot \lesssim m_2 \lesssim 20\,M_\odot$. One has $\epsilon \lesssim 0.25$ for $M \lesssim 30\,M_\odot$. The implied range of binary systems is of clear astrophysical interest. The overestimation shows a strong dependence on both the mass and the asymmetry of systems. 

We note that for systems that are both very massive and very asymmetric, $\epsilon$ can become negative, and $\rho[h] > \rho[h_0]$. This occurs when the dominant harmonic enters the detector's bandwidth in only a small frequency interval. In that case both $\rho[h]$ and $\rho[h_0]$ will be small, at least for initial detectors; as we will show later on, this is not necessarily true for advanced detectors.

\subsection{Analytic treatment}
\label{s:analytic}

We have seen that in initial detectors, modeling signals as restricted waveforms generally leads to an overestimation of signal-to-noise ratio:
\be
\rho[h] < \rho[h_0].
\label{ineq} 
\ee
We now analytically explain (\ref{ineq}) at 1PN order in amplitude and within its domain of validity.

The inequality (\ref{ineq}) is equivalent to $(h|h) < (h_0|h_0)$. Write $h = \sum_{k=1}^7 h^{(k)}$,
where the $h^{(k)}$ are amplitude-corrected harmonics. Leaving out the step functions in frequency, the latter are of the form
\be
\tilde{h}^{(k)}(f) = \mathcal{C} f^{-7/6} 
\left[ \sum_{m=0}^5 \zeta_{(k,m/2)} (2 \pi M f)^{m/3} \right]
\exp\left[i\left(2\pi f t_c -\pi/4 + k\psi\left(\frac{1}{k} f\right)\right)\right],
\label{h2}
\ee
where $\mathcal{C}$ is a real function of chirp mass and distance while the coefficients $\zeta_{(k,m/2)}$ are complex functions of $(\theta,\phi,\psi,\iota,\e)$. (Setting $\zeta_{(k,m/2)} = 0$ for $m > 0$ would lead to the restricted waveform $h_0$.) Now consider waveforms $h$ for which $f_{LSO} \gg f_s$. In that case one can write 
\be
(h|h) \simeq \sum_{k=1}^7 (h^{(k)}|h^{(k)}),
\label{approximation}
\ee
because the various harmonics tend to interfere destructively with each other. (It can be checked numerically that the approximation is valid to within a few percent for stellar mass systems as seen in Initial LIGO.) The contributions $(h^{(k)}|h^{(k)})$ as well as $(h_0|h_0)$ take the form of integrals, the integrands of which all have a common factor $1/S_h(f)$. 
For brevity, let $h$ be of the $(1,3.5)$PN type, so that only the first four harmonics are present. Then up to order $(2\pi M f)^{4/3}$, the remaining functions in the integrand can be written as 
\ba
&&\mbox{Re}[(\tilde{h}^{(1)}(f))^\ast \tilde{h}^{(1)}(f)] +
  \mbox{Re}[(\tilde{h}^{(2)}(f))^\ast \tilde{h}^{(2)}(f)] +
  \mbox{Re}[(\tilde{h}^{(3)}(f))^\ast \tilde{h}^{(3)}(f)] +
  \mbox{Re}[(\tilde{h}^{(4)}(f))^\ast \tilde{h}^{(4)}(f)] \nn\\
&& \quad = \mathcal{C}^2 f^{-7/3} \nn\\
&& \quad\quad \times \left[|\zeta_{(2,0)}|^2 \right.\nn\\
&& \quad\quad\quad\quad \left.
+ (2 \pi M f)^{2/3}\left(|\zeta_{(1,1/2)}|^2 + |\zeta_{(3,1/2)}|^2 + 2 \mbox{Re}[\zeta^\ast_{(2,0)}\zeta_{(2,1)}]\right) \right. \nn\\
&& \quad\quad\quad\quad \left.
+ (2 \pi M f)^{4/3}\left(|\zeta_{(2,1)}|^2+|\zeta_{(4,1)}|^2]\right)\right].
\label{integrands}
\ea
Now,
\be
|\zeta_{(1,1/2)}|^2 + |\zeta_{(3,1/2)}|^2 + 2 \mbox{Re}[\zeta^\ast_{(2,0)}\zeta_{(2,1)}] \leq 0
\label{inequalities}
\ee
for \emph{any} parameter values, and so one should expect $(h|h)<(h_0|h_0)$, whence $\rho[h] < \rho[h_0]$. For an explicit proof we refer to Appendix B.

Obviously the inequality (\ref{ineq}) will not hold for systems where $2 f_{LSO} < f_s < 7 f_{LSO}$, because then $(h_0|h_0)=0$ while $(h|h) > 0$ because of the contribution of the higher harmonics; recall that we have disregarded the step functions $\theta(k f_{LSO} - f)$. Also, if the second harmonic enters the bandwidth in too small a frequency interval then the approximation (\ref{approximation}) breaks down and the inequality (\ref{ineq}) may again be reversed; hence the regions of negative $\epsilon$ in Fig.~\ref{f:SNR_Overestimation}. Both of these points will be important in the discussion of advanced detectors (section \ref{s:advanced} below). Finally, if $h$ had been a $(0.5, 3.5)$PN waveform then in the inequality (\ref{inequalities}) only the strictly positive contributions would have survived, so that once more $\rho[h] > \rho[h_0]$. This is as observed by Sintes and Vecchio in \cite{SV}, and it is also evident in Table \ref{t:PNdependence}.

\section{Advanced detectors}
\label{s:advanced}

We now turn to Advanced LIGO and EGO; for the former we use the estimated power spectral density from \cite{Arunetal} while an analytic fit of the EGO PSD can be found in Appendix C. In addition to stellar mass binary inspirals, these will be able to see inspirals of intermediate-mass black hole binaries with total masses between 50 and more than a thousand solar masses. The latter may be formed in galactic nuclei in the process that leads to the formation of a supermassive black hole or in globular clusters \cite{QuinlanShapiro}. Coalescence rates are uncertain; assuming they are only $10^{-4}$ that of binary neutron stars \cite{FlanaganHughes}, there will be a few events per year within a distance of 2 Gpc.

Below we first briefly discuss the effect of higher harmonics on the mass reach. Next we look at how they affect the redshift to which inspirals can be detected, after which we compute some estimates of detection rates with restricted and amplitude-corrected waveforms.

\subsection{Mass reach}

\begin{figure}[htbp!]
\centering
\includegraphics[scale=0.40,angle=0]{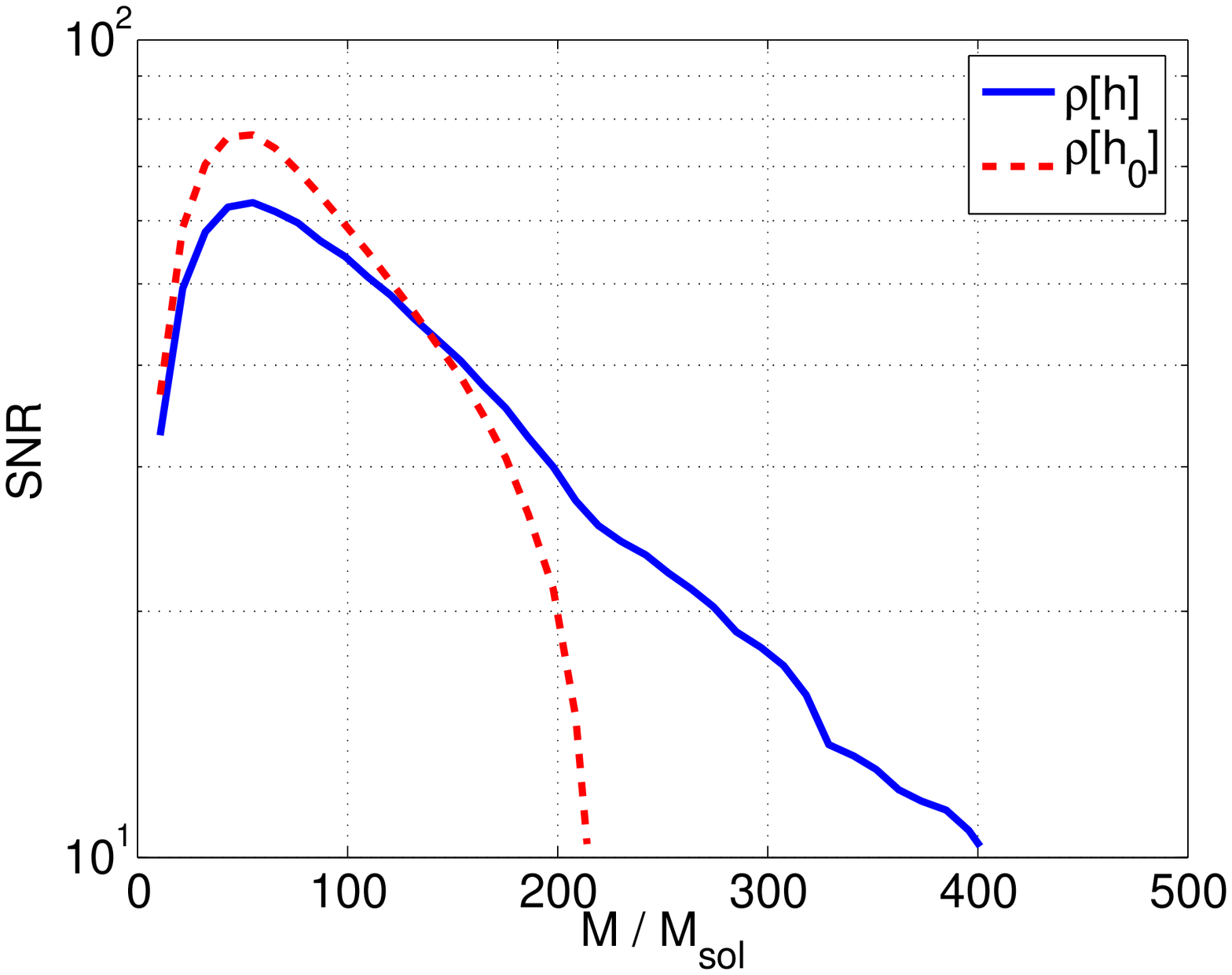}
\includegraphics[scale=0.40,angle=0]{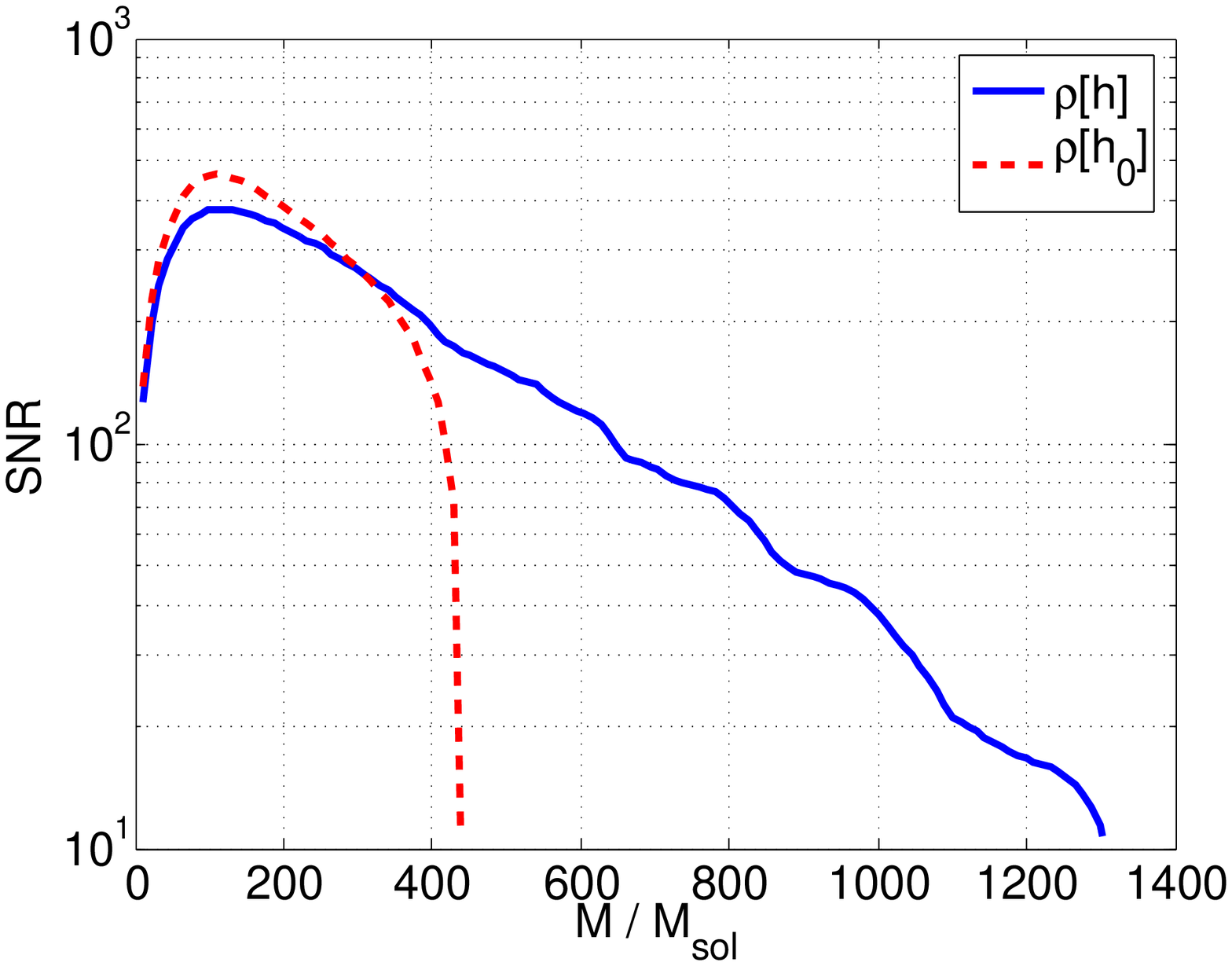}
\caption{Plots of $\rho[h]$ and $\rho[h_0]$ as functions of total mass for Advanced LIGO (left panel) and EGO (right). Distance is fixed at 100 Mpc, and we assume $m_1/m_2 = 0.1$. Angles are as in Table \ref{t:PNdependence}. At low masses one has $2 f_{LSO} \gg f_s$ and $\rho[h_0]$ dominates. For sufficiently high masses, $2 f_{LSO} \leq f_s$, so that the dominant harmonic no longer enters the detector's bandwidth and the SNR for the restricted waveform vanishes. For such masses, higher harmonics in the amplitude-corrected waveform will continue to enter the bandwidth and can lead to significant SNRs. As a result, at the given distance the use of amplitude-corrected waveforms approximately doubles the mass reach of Advanced LIGO and triples that of EGO.}
\label{f:SNR}
\end{figure}  

Here the emphasis will be on inspiral of intermediate mass binaries. It is generally thought that in a given cluster, usually only one intermediate mass black hole (IMBH) will form, which is left to interact with stellar mass compact objects \cite{QuinlanShapiro,Kalogera}. Inspirals involving an IMBH will then tend to be asymmetric, with $m_1/m_2$ roughly in the range $10^{-2}$--$10^{-1}$. In Fig.~\ref{f:SNR} we plot the SNRs $\rho[h]$ and $\rho[h_0]$ against total mass $M$, keeping $m_1/m_2$ fixed at $0.1$ and taking the distance to be 100 Mpc. Here too, for low masses, modeling signals and templates as restricted waveforms would lead to a consistent overestimation of SNR: $\rho[h_0] > \rho[h]$. However, as mass is increased, $\rho[h]$ gradually comes to dominate. Eventually the $k=2$ harmonic no longer reaches the detector's bandwidth, at which point $\rho[h_0]$ drops to zero. This occurs when $2 f_{LSO} = (6^{3/2} \pi M)^{-1} = f_s$, with $f_s$ the detector-dependent lower cut-off frequency. However, for higher masses, the higher harmonics will still be in the bandwidth, and as can be seen from the plots of $\rho[h]$, they can lead to a sizeable SNR. As a result, at a distance of 100 Mpc the use of amplitude-corrected templates approximately doubles the mass range accessible to Advanced LIGO, and it triples that of EGO! 

Let us assume event rates for intermediate mass inspiral to be in the order of a few per year within a distance of 2 Gpc. As can be inferred from Fig.~\ref{f:SNR} (by rescaling SNRs by a factor of 20), at that distance the SNR $\rho[h]$ in Advanced LIGO will stay below 4 for $m_1/m_2 = 0.1$. By contrast, in EGO one has $\rho[h] > 5$ up to $M \sim 600\,M_\odot$ with a maximum of $\rho[h] \sim 20$ for $M \sim 100\,M_\odot$, the approximate lower limit for intermediate-mass binaries. Although such inspirals will be accessible to Advanced LIGO, they would have to happen at distances of at most a few hundred Mpc to be seen with appreciable SNR, and coalescence rates are likely to be low within the corresponding spatial volume. This is much less of a problem for EGO, which has a significantly larger spatial coverage.

\subsection{Redshift reach}

In Fig.~\ref{f:SNR}, distance was fixed at 100 Mpc, leading to large SNRs even when the dominant harmonic does not enter the bandwidth. We may now ask to what distances sources can be seen depending on how many harmonics reach the bandwidth. Since large distances will be involved, a proper treatment will have to take into account the effects of cosmological redshift $z$. This is achieved by making the following replacements in waveforms:
\ba
r \longrightarrow D(z), && \M \longrightarrow \M' = (1+z)\M, \nn\\
t_c \longrightarrow t_c' = (1+z) t_c, && f \longrightarrow f' = \frac{f}{1+z},
\ea
where in a flat Universe the luminosity distance $D$ depends on redshift as
\be
D(z) = \frac{1}{H_0}(1 + z)\int_0^z \frac{dz'}{\sqrt{\Omega_{matter}(1+z')^3 + \Omega_\Lambda}},
\label{luminositydistance}
\ee
with $H_0$ the Hubble parameter at the current epoch while $\Omega_{matter}$ and $\Omega_\Lambda$ are dimensionless quantities defined on the basis of the total matter density and the cosmological constant, respectively. Wherever specific values are needed we will set $H_0 = 71$ km/(Mpc s) and $\Omega_{matter} = 0.27$, $\Omega_\Lambda = 0.73$ \cite{WMAP}.
Note that the redshifting of frequencies will also affect the \emph{observed} frequency of last stable orbit: $f_{LSO}' = f_{LSO}/(1+z)$. In general, the observed quantities are the redshifted ones. Nevertheless, when specifying systems we will continue to do so in terms of \emph{physical} mass. In older work, precedence has usually been given to observed mass. In the present study we will be dealing with relatively large redshifts, and in order to gain a clear understanding of the astrophysical implications of the results (in particular, we will want to distinguish between stellar mass and intermediate mass systems) it is then more convenient to work with physical mass. As an aside we note that assuming knowledge of $H_0$, $\Omega_{matter}$ and $\Omega_\Lambda$, and given a network of (at least three) detectors, luminosity distance can be measured \cite{GurselTinto, Schutz, Schutz1} and $z$ can be solved for by inverting the relation (\ref{luminositydistance}). Together with the observed mass, this leads to a value for the physical mass.

\begin{figure}[htbp!]
\centering
\includegraphics[scale=0.40,angle=0]{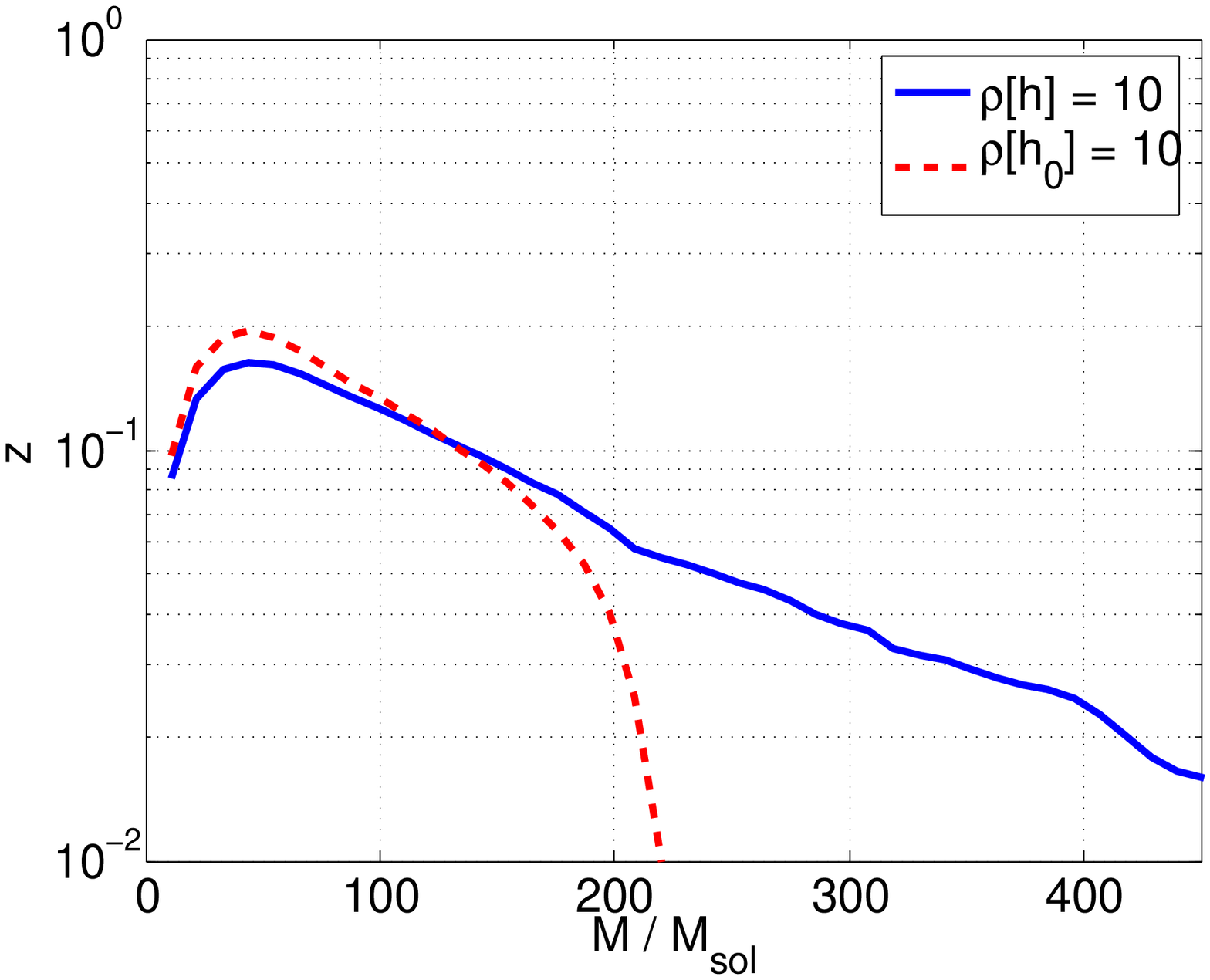}
\includegraphics[scale=0.40,angle=0]{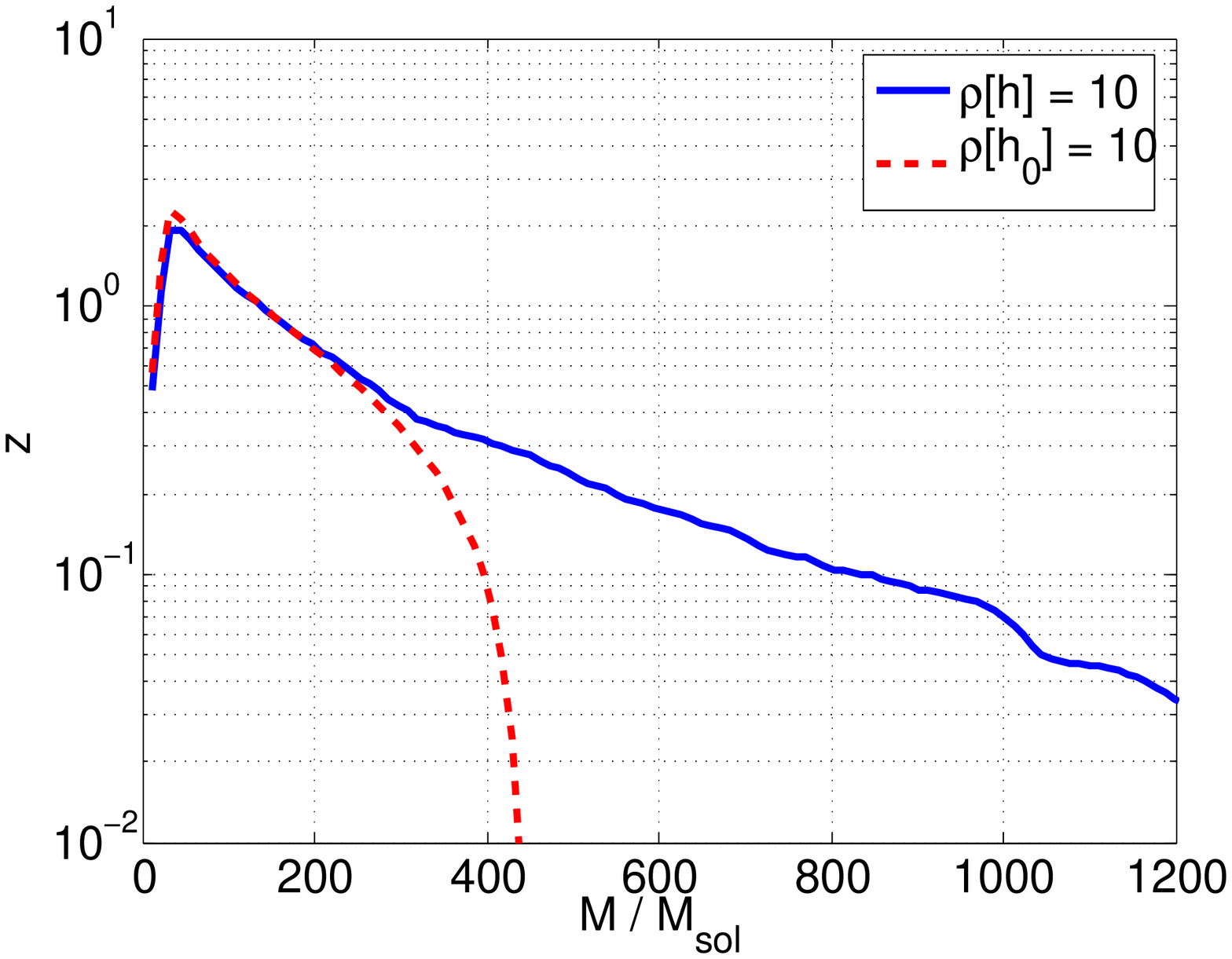}
\caption{The redshift reach of Advanced LIGO (left) and EGO (right) as functions of (physical) total mass for fixed SNRs of 10 with amplitude-corrected and restricted waveforms. We have fixed $m_1/m_2 = 0.1$, and angles are as in Table \ref{t:PNdependence}.}
\label{f:z}
\end{figure}  

\begin{table}[htp!] 
\begin{tabular}{lrcrcrcrcrc} 
\hline 
\hline 
$(\theta,\phi)$ &\vline& $(\iota,\psi)$ &\vline& $z$ (AdvLIGO, $h_0$) &\vline& $z$ (AdvLIGO, $h$) &\vline& $z$ (EGO, $h_0$) &\vline& $z$ (EGO, $h$) \\
\hline
$(0,0)$ &\vline& $(0,0)$ &\vline& 0.248 &\vline& 0.198 &\vline& 2.306 &\vline& 2.157 \\
\hline
$(\pi/6,\pi/3)$ &\vline& $(\pi/6,0)$ &\vline& 0.191 &\vline& 0.160 &\vline& 1.865 &\vline& 1.752 \\
"         &\vline& $(\pi/3,\pi/8)$ &\vline& 0.139 &\vline& 0.130 &\vline& 1.360 &\vline& 1.320 \\
"         &\vline& $(\pi/2,\pi/4)$ &\vline& 0.098 &\vline& 0.097 &\vline& 0.942 &\vline& 0.898 \\
\hline
$(\pi/3,\pi/3)$ &\vline& $(\pi/6,0)$ &\vline& 0.121 &\vline& 0.100 &\vline& 1.176 &\vline& 1.057 \\
"             &\vline& $(\pi/3,\pi/8)$ &\vline& 0.087 &\vline& 0.083 &\vline& 0.829 &\vline& 0.766 \\
"             &\vline& $(\pi/2,\pi/4)$ &\vline& 0.058 &\vline& 0.057 &\vline& 0.524 &\vline& 0.471 \\
\hline 
$(\pi/2,0)$    &\vline& $(\pi/6,0)$ &\vline& 0.114 &\vline& 0.095 &\vline& 1.107 &\vline& 0.992 \\
"             &\vline& $(\pi/3,\pi/8)$ &\vline& 0.075 &\vline& 0.071 &\vline& 0.699 &\vline& 0.641 \\
\hline
\end{tabular}
\caption{Redshift reach given a $(9,90)\,M_\odot$ system in Advanced LIGO and EGO with restricted waveforms $h_0$ and amplitude-corrected waveforms $h$, for $\rho[h_0]=10$ and $\rho[h]=10$. We have again fixed $m_1/m_2 = 0.1$.} 
\label{t:zAngularDependence}
\end{table}

\begin{table}[htp!] 
\begin{tabular}{lrcrcrcrc} 
\hline 
\hline 
$m_1/m_2$ &\vline& $z$ (AdvLIGO, $h_0$) &\vline& $z$ (AdvLIGO, $h$) &\vline& $z$ (EGO, $h_0$) &\vline& $z$ (EGO, $h$) \\
\hline
0.2 &\vline& 0.170 &\vline& 0.154 &\vline& 1.668 &\vline& 1.608 \\
0.4 &\vline& 0.204 &\vline& 0.169 &\vline& 1.965 &\vline& 1.857 \\
0.6 &\vline& 0.217 &\vline& 0.175 &\vline& 2.073 &\vline& 1.942 \\
0.8 &\vline& 0.222 &\vline& 0.178 &\vline& 2.113 &\vline& 1.967 \\
1   &\vline& 0.224 &\vline& 0.179 &\vline& 2.123 &\vline& 1.975 \\
\hline
\end{tabular}
\caption{Redshift reach for different mass ratios given a total mass of $100\,M_\odot$, in Advanced LIGO and EGO with restricted waveforms $h_0$ and amplitude-corrected waveforms $h$, for $\rho[h_0]=10$ and $\rho[h]=10$. Angles are again as in Fig.~\ref{f:z}.} 
\label{t:zMassRatioDependence}
\end{table}

In Fig.~\ref{f:z} we display the redshift at which inspirals can be seen as a function of total mass, still keeping $m_1/m_2 = 0.1$, with a fixed SNR of 10 for the restricted and amplitude-corrected waveforms, in Advanced LIGO and EGO. In the absence of an analytic expression for redshift reach as a function of SNR, redshifts were obtained using an elementary bisection method, producing successively better guesses for $z$ until the value of the SNR differed from the desired one by less than 0.01\%. In the Figure the angles are fixed; angular dependence is explored in Table \ref{t:zAngularDependence}. Finally, Table \ref{t:zMassRatioDependence} explores the dependence of the redshift reach on the ratio of component masses $m_1/m_2$.

First consider Advanced LIGO. Because $\rho[h_0]$ tends to dominate over $\rho[h]$ at low total mass (see Fig.~\ref{f:SNR}), fixing both to be equal to 10 leads to a larger redshift reach for the restricted waveforms at the low mass end. However, the redshift reach for the amplitude-corrected waveforms quickly catches up; after having peaked around $M \sim 50\,M_\odot$ it decreases relatively slowly. A dip is seen just before $M \simeq 220\,M_\odot$, where the dominant harmonic leaves the bandwidth; for the restricted waveform $z$ drops to zero near that mass. From Fig.~\ref{f:z}, Table \ref{t:zAngularDependence}, and Table \ref{t:zMassRatioDependence} we may conclude that with restricted and non-restricted waveforms alike, a large fraction of inspirals with $M \lesssim 100\,M_\odot$ will be detectable at redshifts of $z \sim 0.05$, and several times farther away depending on mass ratio, sky position, and orientation.

With EGO the situation is qualitatively similar, but there the redshift reach is typically an order of magnitude higher than for Advanced LIGO. From Fig.~\ref{f:z}, Table \ref{t:zAngularDependence}, and Table \ref{t:zMassRatioDependence} we infer that a large fraction of inspirals with $M \lesssim 100\,M_\odot$ are detectable with $\rho[h_0]=10$ at redshifts $z \gtrsim 0.5$, although this number can again be several times larger with better mass ratio, sky position, and/or orientation. Well positioned and oriented stellar mass inspirals are visible throughout a significant part of the visible Universe.

\subsection{Detection rates}

Phinney \cite{Phinney} estimated the number of stellar mass inspiral events out to a given distance $d_{max}$ as
\be
R_{insp} = \frac{L(d_{max})}{L_{MW}}R_{MW}, 
\label{Rinsp}
\ee
where $L(d_{max})$ is the accumulated blue band luminosity for distances $d \leq d_{max}$, $L_{MW}$ the blue luminosity of the Milky Way, and $R_{MW}$ the coalescence rate in a Milky Way equivalent galaxy. Tutukov and Yungelson \cite{TutukovYungelson} arrived at the estimate $R_{MW} \simeq 1.4 \times 10^{-6}\,\mbox{yr}^{-1}$ for binary black hole inspirals. From the Lyon-Meudon extragalactic database \cite{Paturel} or the Tully Nearby Galaxy Catalog \cite{Tully} one infers \cite{Nutzmanetal}
\be
\frac{L_{tot}(d_{max})}{L_{MW}} \simeq 10^{-2}\left(\frac{d_{max}}{1\,\mbox{Mpc}}\right)^3
\label{BL}
\ee
out to distances of a few hundred Mpc.

We are interested in obtaining detection rates over cosmological distances. For simplicity we will assume that the event rate per comoving volume is the same everywhere in the Universe.

Consider black hole binaries at redshift $z$, inspiraling at a rate $\dot{N} = 3/(4\pi)\,\times\,10^{-2}\,\mbox{Mpc}^{-3}R_{MW}$ per comoving volume per unit of cosmic time local to the event. As seen from Earth, the event rate between redshifts $z$ and $z + dz$ is \cite{Phinney2}
\be
\frac{\partial^2\mathcal{N}}{\partial t \partial z }\,dz = \frac{\dot{N}}{1+z} \frac{dV_c}{dz}\,dz,
\ee
where the comoving volume element per unit redshift is
\be
\frac{dV_c}{dz} = 4\pi \frac{1}{H_0} \frac{r_c^2(z)}{\sqrt{\Omega_{matter}(1+z)^3 + \Omega_\Lambda}}, 
\ee
with $r_c(z)$ the comoving distance; in terms of luminosity distance one has $r_c(z) = D(z)/(1+z)$.
The total event rate, as seen from Earth, up to redshift $z_{max}$ is then 
\ba
R &=& \int_0^{z_{max}} \frac{\partial^2\mathcal{N}}{\partial t \partial z }\,dz\nn\\
&=& \frac{4\pi \dot{N}}{H_0} \int_0^{z_{max}} \frac{D^2(z)\,dz}{(1+z)^3\sqrt{\Omega_{matter}(1+z)^3 + \Omega_\Lambda}},
\ea
where $D(z)$ is given by (\ref{luminositydistance}).

In the previous subsection we saw that stellar mass binaries will be detectable in Advanced LIGO at redshifts of $z \sim 0.05$ and much beyond. Setting $z_{max} = 0.05$, we arrive at a (conservative) detection rate of 
\be  
R^{AdvL}_{stellar} \simeq 0.1 \,\mbox{yr}^{-1}.
\ee

We have seen that EGO would be able to detect most stellar mass binaries with redshifts $z \gtrsim 0.5$; setting $z_{max}$ equal to this value we arrive at a detection rate 
\be
R^{EGO}_{stellar} \simeq 70 \,\mbox{yr}^{-1},
\ee
corresponding to about one detection every few days.

Following Fig.~\ref{f:z}, for intermediate mass binaries with mass $50\,M_\odot \lesssim M \lesssim 400\,M_\odot$, the amplitude-corrected waveform leads to a redshift reach $z \gtrsim 0.3$, while with the restricted waveform one has $z \gtrsim 0.1$. Let $R^{EGO}_{IM, full}$ be the detection rate from the amplitude-corrected waveform and $R^{EGO}_{IM, restr}$ the one from the restricted waveform. If in each case we set $z_{max}$ equal to above values,
\be 
\frac{R^{EGO}_{IM, full}}{R^{EGO}_{IM, restr}} \simeq 20.
\ee

Needless to say, the above estimates are only indicative (and most likely overly conservative); one could perform a more in-depth analysis along the lines of \cite{Rates,Nutzmanetal}, but that would be outside the scope of this paper. Nevertheless, they already provide dramatic demonstrations of how important it will be to use the amplitude-corrected rather than the restricted waveforms for advanced detectors.

\section{Conclusions and future directions}

We have investigated the effect on signal-to-noise ratio and distance reach of the presence or absence of amplitude corrections in post-Newtonian gravitational waveforms for the quasi-circular, adiabatic inspiral of binary compact objects. This was studied using the stationary phase approximation, which allowed us to gain direct analytic insight. Our conclusions may be summarized as follows.

\begin{itemize}

\item For low-mass systems ($M \lesssim 30\,M_\odot$), which are the ones of main interest for initial interferometric detectors, the use of restricted waveforms as templates and simulated signals can lead to significantly larger signal-to-noise ratios and detection rates compared to what is attained by the best available, $(2.5,3.5)$PN amplitude-corrected waveforms. This runs counter to what one might naively conclude from earlier work \cite{SV}, where only 0.5PN amplitude corrections were taken into account. Indeed, in going from 0PN to 0.5PN in amplitude, SNRs will increase; however, from 1PN onwards we observed a steady decrease in the SNRs $\rho[h]$ computed from the amplitude-corrected waveforms compared to those of the restricted ones, $\rho[h_0]$. The effect is the most pronounced for extreme to moderately asymmetric systems with large masses. For total mass up to $30\,M_\odot$ the difference between $\rho[h]$ and $\rho[h_0]$ can be as large as 25\%, which, in initial detectors, corresponds to a difference of almost a factor of two in detection rates. We studied this effect analytically at the 1PN level and found that, for the specified systems, it results from the amplitude corrections to the second harmonic.

\item Advanced ground-based detectors such as Advanced LIGO and EGO will be more sensitive at low frequencies. As a result, they will (i) be able to detect stellar mass systems at much larger distances; and (ii) be sensitive to signals from more massive inspiraling binaries such as very asymmetric, intermediate-mass systems with total mass $50\,M_\odot \lesssim M \lesssim 1000\,M_\odot$. In the time domain it is natural to terminate waveforms at a time determined by $F(t) = f_{LSO}$, where $f_{LSO}$ is the orbital frequency at the last stable orbit. In the SPA this will roughly correspond to cutting off the $k$th harmonic at $f = k f_{LSO}$. Thus, higher harmonics might still enter the bandwidth even if the leading-order, $k = 2$, harmonic does not, and the resulting SNRs can be considerable. The consequences are twofold. At relatively small distances ($\sim 100$ Mpc), the detector's mass reach will increase significantly, by about a factor of two in Advanced LIGO and a factor of three in EGO. This would allow for the detection of intermediate-mass inspirals with much higher total mass than with the restricted waveforms. Secondly, over cosmological distances, restricted waveforms underestimate detection rates of intermediate-mass inspirals by at least a factor of 20.

\end{itemize}

Our results can be extended in various directions.

\begin{itemize}

\item \emph{Comparison with other waveforms.} We have investigated how simulations involving amplitude-corrected waveforms both for templates and simulated signals would compare with simulations involving only restricted templates and ``signals". Since the $(2.5,3.5)$PN waveforms are presumably the closest approximations we have to real signals, it would be interesting to see how effective the standard Taylor, EOB, Pad\'e, and BCV waveforms are in detecting them.

\item \emph{Parameter estimation.} Some consequences of other harmonics for parameter estimation were studied by Hellings and Moore \cite{HellingsMoore} for the case of LISA. A systematic treatment indicating qualitative differences in parameter estimation with restricted and non-restricted waveforms for ground-based detectors will appear in \cite{paramest}, and a similar study for LISA is in preparation. 

\item \emph{Testing general relativity.} The post-Newtonian phasing of alternative theories of gravity (in particular scalar-tensor and massive graviton theories) has been worked out to 1PN order by Berti, Buonanno and Will \cite{BBW}, who also studied the accuracy with which these could be distinguished in the context of LISA. Another recent proposal by Arun et al.~\cite{Testing} exploits the interdependence of the parameters $\psi_i$ in the phasing of waveforms (see Appendix A) to look for deviations from general relativity. It would be of great interest to devise similar tests that also take amplitude corrections into account.

\end{itemize}

\section*{Acknowledgements}

It is a pleasure to thank S.~Babak and B.S.~Sathyaprakash for very useful discussions. We are especially grateful to M.~Punturo for sharing with us the power spectral density for EGO. This research was supported by PPARC grant PP/B500731/1.

\section*{Appendix A: Structure of the amplitude-corrected waveforms in the stationary phase approximation}

The expression for the phase is (\ref{phase}), where the coefficients $\psi_i$ are given by\footnote{We remind the reader that our conventions are slightly different from what one often finds in other papers, as explained in Section \ref{s:waveforms}. Frequency dependence in equations (\ref{phase}), (\ref{psicoefficients}), and (\ref{FdotF}) is to be read with this caveat in mind.} \cite{2.5PN,3.5PN,Arunetal}
\ba
\psi_0 &=& 1,\nn\\
\psi_1 &=& 0,\nn\\
\psi_2 &=& \frac{20}{9}\,\left[\frac{743}{336} + \frac{11}{4}\eta\right],\nn\\
\psi_3 &=& -16\pi,\nn\\
\psi_4 &=& 10\,\left[\frac{3058673}{1016064} + \frac{5429}{1008}\eta + \frac{617}{144}\eta^2 \right],\nn\\
\psi_5 &=& \pi\,\left[\frac{38645}{756} + \frac{38645}{756}\ln\left(\frac{f}{f_{LSO}}\right) - \frac{65}{9}\eta\left(1 + \ln\left(\frac{f}{f_{LSO}}\right)\right)\right],\nn\\
\psi_6 &=& \left(\frac{11583231236531}{4694215680} - \frac{640\pi^2}{3} - \frac{6848\gamma}{21}\right) \nn\\
&& + \e\,\left(-\frac{15335597827}{3048192} + \frac{2255\pi^2}{12} - \frac{1760\theta}{3} + \frac{12320\lambda}{9}\right) \nn\\
&& + \frac{76055}{1728}\e^2 - \frac{127825}{1296}\e^3 - \frac{6848}{21}\ln\left[4(2\pi M f)^{1/3}\right], \nn\\
\psi_7 &=& \pi\left(\frac{77096675}{254016} + \frac{378515}{1512}\e-\frac{74045}{756}\e^2\right),\nn\\  
\label{psicoefficients}
\ea
with $\gamma=0.5772 \dots$ the Euler-Mascheroni constant. The parameters $\lambda$ and $\theta$ were unknown until recently because of regularization ambiguities; they have been determined to be $\lambda=-\frac{1987}{3080}$ and $\theta=-\frac{11831}{9240}$ \cite{3.5PN}, completing the phasing formula up to 3.5PN.

In the expressions (\ref{h_k}) for the $\tilde{h}^{(k)}$ we take the ``frequency sweep" $\dot{F}$ to be
\ba
\dot{F} &=& \frac{48}{5\pi\M^2}(2\pi \M F)^{11/3} [1 - \left(\frac{743}{336}+\frac{11}{4}\eta\right)(2\pi M F)^{2/3} + 4\pi (2\pi M F) \nn\\
&& \quad + \left(\frac{34103}{18144}+\frac{13661}{2016}\eta + \frac{59}{18}\eta^2 \right)(2\pi M F)^{4/3} - \left(\frac{4159\pi}{672} + \frac{189\pi}{8}\eta\right)(2\pi M F)^{5/3}],\nn\\
\label{FdotF}
\ea
where $\M=M\e^{3/5}$ is the chirp mass. 
To 2.5PN order,
\ba
&&\frac{1}{\sqrt{\dot{F}(t(f))}} \nn\\ 
&& \quad = \sqrt{\frac{5\pi}{48}}\frac{\M}{(2\pi\M f)^{11/6}} \left[1 + S_1 (2\pi M f)^{2/3} + S_{3/2} (2\pi M f) + S_2 (2\pi M f)^{4/3} + S_{5/2} (2\pi M f)^{5/3}\right] \nn\\
\ea
where
\ba
S_1 &=& \frac{1}{2} \left(\frac{743}{336} + \frac{11}{4}\e\right), \nn\\
S_{3/2} &=& -2\pi, \nn\\
S_2 &=& \frac{7266251}{8128512} + \frac{18913}{16128}\e + \frac{1379}{1152}\e^2, \nn\\
S_{5/2} &=& -\pi \frac{4757}{1344} - \frac{3}{16}(-63+44\pi)\e.\nn\\
\label{freqsweepcoefficients}
\ea
The coefficients $H^{(s)}_{+,\times}$ in (\ref{hpluscross}) can be found in \cite{2.5PN}, and we will not repeat them here. They are linear combinations of sines and cosines of multiples of the orbital phase $\Psi$. The prefactors in these expressions will be denoted $C_{+,\times}^{(n,s)}$ and $D_{+,\times}^{(n,s)}$, the former being the prefactors of $\cos(n\Psi)$ and the latter the prefactors of $\sin(n\Psi)$. Now define\footnote{Up to and including 1.5PN order, the $H^{(s)}_+$ involve cosines only and the $H^{(s)}_\times$ involve sines only, so that $D_+^{(n,s)} = 0$ and $C_\times^{(n,s)} = 0$ until 2PN order.}
\be
P_{(n,s)} \equiv \mbox{sign}\left(C_+^{(n,s)} F_+ + C_\times^{(n,s)} F_\times\right) \sqrt{\left(C_+^{(n,s)} F_+ + C_\times^{(n,s)} F_\times\right)^2 + \left(D_+^{(n,s)} F_+ + D_\times^{(n,s)} F_\times\right)^2},
\ee
and
\be
\varphi_{(n,s)} \equiv \tan^{-1}\left(-\frac{D_+^{(n,s)} F_+ + D_\times^{(n,s)} F_\times}{C_+^{(n,s)} F_+ + C_\times^{(n,s)} F_\times}\right).
\ee

The harmonics contributing to the waveform up to 2.5PN have the following structure.
\ba
\tilde{h}^{(1)}(f) &=&  \frac{\M^{5/6}}{r} \sqrt{\frac{5}{48}}\pi^{-2/3}(2f)^{-7/6}
\left[e^{-i\varphi_{(1,1/2)}}P_{(1,1/2)}(2\pi M f)^{1/3}\right. \nn\\
&& \left.+(e^{-i\varphi_{(1,3/2)}}P_{(1,3/2)} + e^{-i\varphi_{(1,1/2)}}P_{(1,1/2)} S_1)(2\pi M f)\right. \nn\\
&& \left.+(e^{-i\varphi_{(1,2)}}P_{(1,2)} + e^{-i\varphi_{(1,1/2)}}P_{(1,1/2)} S_{3/2})(2\pi M f)^{4/3} \right.\nn\\
&& \left.+(e^{-i\varphi_{(1,5/2)}}P_{(1,5/2)} + e^{-i\varphi_{(1,3/2)}}P_{(1,3/2)} S_1
+ e^{-i\varphi_{(1,1/2)}}P_{(1,1/2)} S_2)(2\pi M f)^{5/3}\right] \nn\\
&&\times \,\theta(f_{LSO} - f)\,\exp\left[i\left(2\pi f t_c - \pi/4 + \psi(f)\right)\right],
\ea
\ba
\tilde{h}^{(2)}(f) &=& 2^{-1/2}\frac{\M^{5/6}}{r} \sqrt{\frac{5}{48}}\pi^{-2/3} f^{-7/6}
\left[e^{-i\varphi_{(2,0)}}P_{(2,0)} \right. \nn\\
&& \left.+ (e^{-i\varphi_{(2,1)}}P_{(2,1)} + e^{-i\varphi_{(2,0)}}P_{(2,0)} S_1)(\pi M f)^{2/3} \right.\nn\\
&& \left.+ (e^{-i\varphi_{(2,3/2)}}P_{(2,3/2)} + e^{-i\varphi_{(2,0)}}P_{(2,0)} S_{3/2})(\pi M f) \right.\nn\\
&& \left.+(e^{-i\varphi_{(2,2)}}P_{(2,2)} + e^{-i\varphi_{(2,1)}}P_{(2,1)}S_1 + e^{-i\varphi_{(2,0)}}P_{(2,0)}S_2)(\pi M f)^{4/3} \right.\nn\\
&& \left.+(e^{-i\varphi_{(2,5/2)}}P_{(2,5/2)} + e^{-i\varphi_{(2,3/2)}}P_{(2,3/2)} S_1 + e^{-i\varphi_{(2,1)}}P_{(2,1)} S_{3/2} + e^{-i\varphi_{(2,0)}}P_{(2,0)} S_{5/2})(\pi M f)^{5/3}\right] \nn\\
&&\times \,\theta(2 f_{LSO} - f)\,\exp\left[i\left(2\pi f t_c - \pi/4 + 2\psi(f/2)\right)\right],
\ea
\ba
\tilde{h}^{(3)}(f) &=& 3^{-1/2}\frac{\M^{5/6}}{r} \sqrt{\frac{5}{48}}\pi^{-2/3}(2f/3)^{-7/6}
\left[e^{-i\varphi_{(3,1/2)}}P_{(3,1/2)} (2\pi M f/3)^{1/3}\right.  \nn\\
&& \left.+ (e^{-i\varphi_{(3,3/2)}}P_{(3,3/2)} + e^{-i\varphi_{(3,1/2)}}P_{(3,1/2)} S_1)(2\pi M f/3) \right.\nn\\
&& \left.+ (e^{-i\varphi_{(3,2)}}P_{(3,2)} + e^{-i\varphi_{(3,1/2)}}P_{(3,1/2)} S_{3/2})(2\pi M f/3)^{4/3}\right. \nn\\
&& \left.+ (e^{-i\varphi_{(3,3/2)}}P_{(3,3/2)} S_1 + e^{-i\varphi_{(3,1/2)}}P_{(3,1/2)} S_2)(2\pi M f/3)^{5/3}\right]\nn\\
&& \times \,\theta(3 f_{LSO} - f)\,\exp[i(2\pi f t_c - \pi/4 + 3\psi(f/3))],
\ea
\ba
\tilde{h}^{(4)}(f) &=& 4^{-1/2}\frac{\M^{5/6}}{r} \sqrt{\frac{5}{48}}\pi^{-2/3} (f/2)^{-7/6}
\left[e^{-i\varphi_{(4,1)}}P_{(4,1)}(\pi M f/2)^{2/3} \right.\nn\\
&& \left.+ (e^{-i\varphi_{(4,2)}}P_{(4,2)} + e^{-i\varphi_{(4,1)}}P_{(4,1)} S_1)(\pi M f/2)^{4/3} \right.\nn\\
&& \left.+ (e^{-i\varphi_{(4,5/2)}}P_{(4,5/2)} + e^{-i\varphi_{(4,1)}}P_{(4,1)} S_{3/2})(\pi M f/2)^{5/3}\right] \nn\\
&&\times \,\theta(4 f_{LSO} - f)\,\exp[i(2\pi f t_c -\pi/4 + 4 \psi(f/4))],
\ea
\ba
\tilde{h}^{(5)}(f) &=& 5^{-1/2}\frac{\M^{5/6}}{r} \sqrt{\frac{5}{48}}\pi^{-2/3} (2f/5)^{-7/6}
\left[e^{-i\varphi_{(5,3/2)}}P_{(5,3/2)}(2\pi M f/5)\right. \nn\\
&& \left.+ (e^{-i\varphi_{(5,5/2)}}P_{(5,5/2)} + e^{-i\varphi_{(5,3/2)}}P_{(5,3/2)} S_1)(2\pi M f/5)^{5/3}\right] \nn\\
&&\times \,\theta(5 f_{LSO} - f)\,\exp[i(2\pi f t_c - \pi/4 + 5\psi(f/5))], 
\ea
\ba
\tilde{h}^{(6)}(f) &=& 6^{-1/2}\frac{\M^{5/6}}{r} \sqrt{\frac{5}{48}}\pi^{-2/3} (f/3)^{-7/6}
e^{-i\varphi_{(6,2)}}P_{(6,2)}(\pi M f/3) \nn\\
&& \times \,\theta(6 f_{LSO} - f)\,\exp[i(2\pi f t_c -\pi/4 + 6\psi(f/6))], 
\ea
\ba
\tilde{h}^{(7)}(f) &=& 7^{-1/2}\frac{\M^{5/6}}{r} \sqrt{\frac{5}{48}}\pi^{-2/3} (2f/7)^{-7/6}
e^{-i\varphi_{(7,7)}}P_{(7,7)}(2\pi M f/7)^{5/3}\nn\\
&&\times \,\theta(7 f_{LSO} - f)\,\exp[i(2\pi f t_c - \pi/4 + 7\psi(f/7))].
\ea
These can in turn be written as
\ba
\tilde{h}^{(k)}(f) &=&  \frac{\M^{5/6}}{r} \sqrt{\frac{5}{48}}\pi^{-2/3} f^{-7/6}\theta(k f_{LSO} - f)\, 
\left[ \sum_{m=0}^5 \zeta_{(k,m/2)} (2 \pi M f)^{m/3} \right] \nn\\
&& \quad\quad\quad\quad\quad \times \exp\left[i\left(2\pi f t_c -\pi/4 + k\psi\left(\frac{1}{k} f\right)\right)\right].
\ea
In the main text we need the $\zeta_{(k,m/2)}$ for a $(1,3.5)$PN waveform. The non-zero coefficients appearing in such a waveform are given by (with $c=\cos(\iota)$ and $s=\sin(\iota)$):
\ba
\zeta_{(1,1/2)} &=& 2^{-7/6} e^{-i\varphi_{(1,1/2)}}P_{(1,1/2)} \nn\\
&=& 2^{-7/6}\mbox{sign}\left[-\frac{s}{8}\sqrt{1-4\eta}\,(5+c^2)F_+\right]\nn\\
&&\quad \times \sqrt{\frac{s^2}{64}(1-4\eta)(5+c^2)^2 F_+^2
+ \frac{9}{16}s^2c^2(1-4\eta) F_\times^2} \nn\\
&&\quad \times \exp\left[-i\tan^{-1}\left(\frac{-6cF_\times}{(5+c^2)F_+}\right)\right],
\nn\\
\label{C11/2}
\ea
\ba
\zeta_{(2,0)} &=& 2^{-1/2} e^{-i\varphi_{(2,0)}}P_{(2,0)} \nn\\
&=& 2^{-1/2}\mbox{sign}\left[-(1+c^2) F_+\right] \nn\\
&& \times \sqrt{(1+c^2)^2 F_+^2 + 4c^2F_\times^2} \nn\\
&& \times \exp\left[-i\tan^{-1}\left(\frac{-2cF_\times}{(1+c^2)F_+}\right)\right], \nn\\
\label{C20}
\ea
\ba
\zeta_{(2,1)} &=& 2^{-1/2} 2^{-2/3} \left[e^{-i\varphi_{(2,1)}}P_{(2,1)} + e^{-i\varphi_{(2,0)}}P_{(2,0)}S_1 \right] \nn\\
&=& 2^{-7/6}\frac{1}{6}\mbox{sign}\left[\left(19+9 c^2 -2c^4 -(19 -11c^2 -6c^4)\e\right)F_+\right] \nn\\
&& \quad \times \sqrt{\left(19+9 c^2 -2c^4 -(19 -11c^2 -6c^4)\e\right)^2F_+^2 + 4c^2\left(17 - 4c^2 -(13-12c^2)\e\right)^2 F_\times^2} \nn\\
&&\quad \times \exp\left[-i\tan^{-1}\left(-\frac{2c[17-4c^2-(13-12c^2)\e]F_+}{[19+9c^2-2c^4-(19-11c^2-6c^4)\e]F_\times}\right)\right] \nn\\
&& + 2^{-7/6}\frac{1}{2}\left(\frac{743}{336} + \frac{11}{4}\e\right)\,\mbox{sign}\left[-(1+c^2)F_+\right]\,\sqrt{(1+c^2)^2 F_+^2 + 4c^2 F_\times^2} \nn\\
&&\quad \times \exp\left[-i\tan^{-1}\left(\frac{-2cF_\times}{(1+c^2)F_+}\right)\right],\nn\\
\label{C21}
\ea
\ba
\zeta_{(3,1/2)} &=& 3^{-1/2}\left(\frac{2}{3}\right)^{-7/6} 3^{-1/3}
e^{-i\varphi_{(3,1/2)}}P_{(3,1/2)} \nn\\
&=& 2^{-7/6} 3^{1/3} \mbox{sign}\left[+\frac{9}{8} s \sqrt{1-4\eta}\,(1+c^2) F_+\right]\nn\\
&&\quad \times \sqrt{\frac{81}{64}s^2(1-4\eta)(1+c^2)^2 F_+^2
+ \frac{81}{16}s^2c^2(1-4\eta) F_\times^2} \nn\\
&&\quad \times \exp\left[-i\tan^{-1}\left(\frac{-2cF_\times}{(1+c^2)F_+}\right)\right],
\nn\\
\label{C31/2}
\ea
\ba
\zeta_{(4,1)} &=& 4^{-1/2}\left(\frac{1}{2}\right)^{-7/6} 4^{-2/3}
e^{-i\varphi_{(4,1)}}P_{(4,1)} \nn\\
&=& 2^{-7/6} \mbox{sign}\left[-\frac{4}{3}s^2(1+c^2)(1-3\eta) F_+\right]\nn\\
&&\quad \times \sqrt{\frac{16}{9}s^4(1+c^2)^2(1-3\eta)^2F_+^2 + \frac{64}{9}s^4c^2(1-3\eta)^2F_\times^2}\nn\\
&&\quad \times \exp\left[i\frac{-2cF_\times}{(1+c^2)F_+}\right].\nn\\
\label{C41}
\ea

\section*{Appendix B: Proof of the inequality (\ref{inequalities})}

From the expressions (\ref{C11/2}), (\ref{C20}), (\ref{C21}), and (\ref{C31/2}) for $\zeta_{(1,1/2)}$, $\zeta_{(2,0)}$, $\zeta_{(2,1)}$, and $\zeta_{(3,1/2)}$, one has
\ba
&&|\zeta_{(1,1/2)}|^2 + |\zeta_{(3,1/2)}|^2 + 2\mbox{Re}[\zeta^\ast_{(2,0)}\zeta_{(2,1)}] \nn\\
&& =  2^{-7/3}\left[\frac{s^2}{64}(1-4\eta)(5+c^2)^2 F_+^2 + \frac{9}{16}s^2c^2(1-4\eta) F_\times^2\right] \nn\\
&& \quad + 2^{-7/3} 3^{2/3}\left[\frac{81}{64}s^2(1-4\eta)(1+c^2)^2 F_+^2 + \frac{81}{16}s^2c^2(1-4\eta) F_\times^2\right] \nn\\
&& \quad - \frac{2^{-2/3}}{6} \cos\left[\tan^{-1}\left(\frac{-2cF_\times}{(1+c^2)F_+}\right)-\tan^{-1}\left(-\frac{2c[17-4c^2-(13-12c^2)\e]F_+}{[19+9c^2-2c^4-(19-11c^2-6c^4)\e]F_\times}\right)\right]\nn\\
&& \quad\quad \times \sqrt{(1+c^2)^2 F_+^2 + 4c^2 F_\times^2} \nn\\
&& \quad\quad \times \sqrt{\left(19+9c^2 -2c^4 -(19 -11c^2 -6c^4)\e\right)^2F_+^2 + 4c^2\left(17 - 4c^2 -(13-12c^2)\e\right)^2 F_\times^2}\nn\\
&& \quad + 2^{-5/3}\left(\frac{743}{336} + \frac{11}{4}\e\right)\,\left[(1+c^2)^2 F_+^2 + 4c^2 F_\times^2\right], 
\label{label}
\ea
where we have used that
\be
\mbox{sign}\left[-(1+c^2)F_+\right]\,\mbox{sign}\left[\left(19+9 c^2 -2c^4 -(19 -11c^2 -6c^4)\e\right)F_+\right] = -1.
\ee
Applying basic trigonometric identities in the third term of (\ref{label}),
\ba
&&|\zeta_{(1,1/2)}|^2 + |\zeta_{(3,1/2)}|^2 + 2\mbox{Re}[\zeta^\ast_{(2,0)}\zeta_{(2,1)}] \nn\\
&& =  2^{-7/3}\left[\frac{s^2}{64}(1-4\eta)(5+c^2)^2 F_+^2 + \frac{9}{16}s^2c^2(1-4\eta) F_\times^2\right] \nn\\
&& \quad + 2^{-7/3} 3^{2/3}\left[\frac{81}{64}s^2(1-4\eta)(1+c^2)^2 F_+^2 + \frac{81}{16}s^2c^2(1-4\eta) F_\times^2\right] \nn\\
&& \quad - \frac{2^{-2/3}}{6}\left[(1+c^2)(19+9c^2 -2c^4 -(19 -11c^2 -6c^4)\e)F_+^2 \right. \nn\\
&& \quad\quad\quad\quad\quad \left. + 4c^2(17 - 4c^2 -(13-12c^2)\e)F_\times^2\right] \nn\\
&& \quad + 2^{-5/3}\left(\frac{743}{336} + \frac{11}{4}\e\right)\,\left[(1+c^2)^2 F_+^2 + 4c^2 F_\times^2\right].
\label{expression}
\ea
The RHS can be written as a linear combination of $F_+^2$ and $F_\times^2$, and sufficient conditions for (\ref{expression}) to be negative definite are that the coefficients in this combination are negative for any $c \in [-1,1]$ and $\eta \in [0,1/4]$. It is straightforward but tedious to show that this is the case. Note the presence of the prefactor $(1-4\eta)$ in several of the positive contributions to these coefficients; this explains the strong $\eta$-dependence of the SNR reduction. 

The origin of the inequality (\ref{inequalities}), and similar inequalities one would encounter at higher order, can be understood as follows. The cross-term in the LHS of (\ref{expression}) is of the form $\mbox{Re}[\zeta^\ast_{(k,(m-1)/2)}\zeta_{(k,(m+1)/2)}]$, which has a large (in absolute value) contribution that looks like 
\ba
&&\mathcal{N}\,\mbox{sign}[C_+^{(k,(m-1)/2)}]\,\mbox{sign}[C_+^{(k,(m+1)/2)}]\, \left|P_{(k,(m-1)/2)} P_{(k,(m+1)/2)}\right| \nn\\ 
&&\quad\quad\quad\quad\quad\quad\times \cos(\varphi_{(k,(m-1)/2)} - \varphi_{(k,(m+1)/2)})),
\label{negativedefinite}
\ea
where we have used the notation of Appendix A. Here $\mathcal{N}$ is some positive numerical prefactor. It can be checked that the angles $\varphi_{(k,(m \pm 1)/2)}$ always have the same sign and, being defined through $\tan^{-1}$, they are in the interval $[-\pi/2,\pi/2]$; hence the cosine will be positive. However, $C_+^{(k,(m-1)/2)}$ and $C_+^{(k,(m+1)/2)}$ have opposite signs, so that the expression (\ref{negativedefinite}) is negative definite.

\section*{Appendix C: Specifications and projected noise power spectral density for EGO}

\begin{figure}[htbp!] 
\includegraphics[scale=0.60,angle=0]{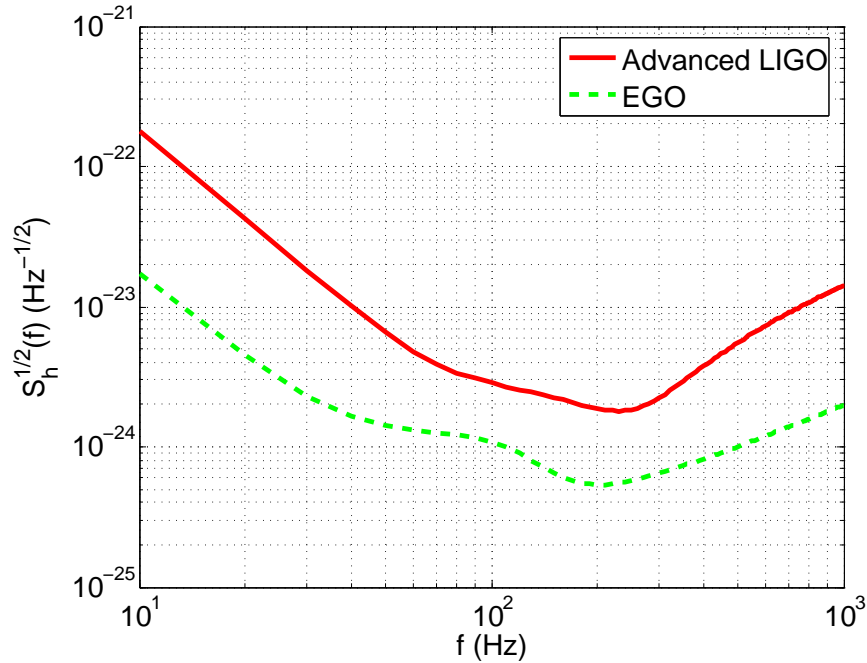}
\caption{The strain sensitivities of Advanced LIGO and EGO as functions of frequency.}
\label{f:PSDs}
\end{figure}

The EGO detector is not yet on the drawing boards. Rather, its noise power spectral density should be viewed as a summary of what is projected to be possible with steady advances in interferometer technology over the next decade or so. For concreteness, a 3 km arm length was assumed, to be placed underground. The optics are kept at cryogenic temperatures ($\sim 5$ K). The loss angle of the mirror substrate and the suspension material is approximately $10^{-9}$. The injected laser power is 100 W, and the recycling factor of the recycling mirror is 50; the finesse of the main cavities is about 600. The radiation pressure noise is suppressed using a scheme suggested by Courty et al.~\cite{Courtyetal}, in which a control cavity is coupled to each mirror, storing a lower light power. The seismic filtering is realized by combining a passive system similar to VIRGO's ``super-attenuator" with an active system as in Advanced LIGO. Above frequencies of a few Hertz the resulting PSD is well-approximated by
\be
S_h(f) = S_0\,\left[x^{p_1} + a_1 x^{p_2} + a_2 \frac{1+b_1 x+b_2 x^2 + b_3 x^3 + b_4 x^4 + b_5 x^5 + b_6 x^6}{1 + c_1 x + c_2 x^2 + c_3 x^3 + c_4 x^4}\right]
\ee
where $x = f/f_0$ with $f_0=200$ Hz, and $S_0 = 1.61 \times 10^{-51}\,\mbox{Hz}^{-1}$. One has
\ba
p_1 = -4.05, &\quad& p_2 = -0.69,  \nn\\
a_1 = 185.62, &\quad& a_2 = 232.56, \nn\\
b_1 = 31.18, \quad b_2 = -64.72, \quad b_3 = 52.24, &\quad& b_4 = -42.16, \quad b_5 = 10.17, \quad b_6 = 11.53, \nn\\
c_1 = 13.58, \quad c_2 = -36.46,  &\quad& c_3 = 18.56, \quad c_4 = 27.43. \nn\\
\ea
The strain sensitivities for Advanced LIGO and EGO are plotted in Fig.~\ref{f:PSDs}.

Finally, the lower cut-off frequencies $f_s$ for EGO as well as Initial and Advanced LIGO are chosen in such a way that for any (untruncated) restricted PN waveform $h_0$, 
\be 
\left[\frac{\int_{f_s}^\infty |h_0(f)|^2 df/S_h(f)}{\int_0^\infty |h_0(f)|^2 df/S_h(f)}\right]^{1/2} 
= \left[\frac{\int_{f_s}^\infty f^{-7/3} df/S_h(f)}{\int_0^\infty f^{-7/3} df/S_h(f)}\right]^{1/2} > 0.99. 
\ee
Picking the largest value of $f_s$ satisfying this requirement and rounding off yields the usual $f_s = 40$ Hz for Initial LIGO and $f_s = 20$ Hz for Advanced LIGO, while $f_s = 10$ Hz for EGO.

\begin{thebibliography}{99}


\bibitem{Observatories} http://www.ligo.caltech.edu/; http://www.virgo.infn.it/;
http://www.geo600.uni-hannover.de/; http://tamago.mtk.nao.ac.jp/

\bibitem{Grishchuketal} L.P.~Grishchuk, V.M.~Lipunov, K.A.~Postnov, M.E.~Prokhorov, and B.S.~Sathyaprakash,
Phys.~Usp.~{\bf 44}, 1-51 (2001);\\
L.P.~Grishchuk, V.M.~Lipunov, K.A.~Postnov, M.E.~Prokhorov, and B.S.~Sathyaprakash,
Usp.~Fiz.~Nauk.~{\bf 171}, 3-59 (2001);\\
L.P.~Grishchuk,
in \emph{Astrophysics Update}, ed.~J.W.~Mason (Springer-Praxis, Berlin, 2004)

\bibitem{Blanchet} L.~Blanchet, 
Liv.~Rev.~Rel.~{\bf 5}, 3 (2002)

\bibitem{2.5PN} 
L.~Blanchet, B.R.~Iyer, C.M.~Will, and A.G.~Wiseman,
Class.~Quantum Grav.~{\bf 13}, 575-584 (1996);\\
K.G.~Arun, L.~Blanchet, B.R.~Iyer, and M.S.S.~Qusailah,
Class.~Quantum Grav.~{\bf 21}, 3771-3802 (2004);
Erratum ibid.~{\bf 22}, 3115 (2005) 

\bibitem{3.5PN}
T.~Damour, P.~Jaranowski, and G.~Sch\"afer,
Phys.~Lett. B {\bf 513}, 147 (2001);\\
Y.~Itoh, T.~Futamase, and H.~Asada,
Phys.~Rev.~D {\bf 63}, 064038 (2001);\\
L.~Blanchet, G.~Faye, B.R.~Iyer, and B.~Joguet,
Phys.~Rev.~D {\bf 65}, 061501(R) (2002);
Erratum ibid.~D {\bf 71}, 129902 (2005);\\
Y.~Itoh and T.~Futamase, 
Phys.~Rev.~D {\bf 68}, 121501(R) (2003);\\
Y.~Itoh, 
Phys.~Rev.~D {\bf 69}, 064018 (2004);\\
L.~Blanchet, T.~Damour, and G.~Esposito-Far\`ese,
Phys.~Rev.~D {\bf 69}, 124007 (2004);\\
L.~Blanchet, T.~Damour, G.~Esposito-Far\`ese, and B.R.~Iyer,
Phys.~Rev.~Lett.~{\bf 93}, 091101 (2004);\\
Y.~Itoh, 
Class.~Quantum Grav.~{\bf 21}, S529-S534 (2004);\\
L.~Blanchet and B.R.~Iyer,
Phys.~Rev.~D {\bf 71}, 024004 (2005)

\bibitem{last3minutes} C.~Cutler et al.,
Phys.~Rev.~Lett.~{\bf 70}, 2984-2987 (1993)

\bibitem{Helstrom} C.W.~Helstrom, \emph{Statistical Theory of Signal Detection} (Pergamon Press,
Cambridge, England, 1968);\\
S.~Babak, R.~Balasubramanian, D.~Churches, T.~Cokelaer, and B.S.~Sathyaprakash, gr-qc/0604037;\\
S.~Babak, T.~Cokelaer, B.S.~Sathyaprakash, and A.S.~Sengupta, \emph{in preparation}

\bibitem{comparison} 
T.~Damour, B.R.~Iyer, and B.S.~Sathyaprakash, 
Phys.~Rev.~D {\bf 63}, 044023 (2001); Erratum ibid.~D {\bf 72}, 029902 (2005)

\bibitem{EOB} 
A.~Buonanno and T.~Damour, 
Phys.~Rev.~D {\bf 59}, 084006 (1999);\\
A.~Buonanno and T.~Damour,
Phys.~Rev.~D {\bf 62}, 064015 (2000)

\bibitem{BCV} 
A.~Buonanno, Y.~Chen and M.~Vallisneri,
Phys.~Rev.~D {\bf 67}, 024016 (2003);\\
A.~Buonanno, Y.~Chen, and M.~Vallisneri,
Phys.~Rev.~D {\bf 67}, 104025 (2003);\\
Y.~Pan, A.~Buonanno, Y.~Chen, and M.~Vallisneri,
Phys.~Rev.~D {\bf 69}, 104017 (2004);\\
A.~Buonanno, Y.~Chen, Y.~Pan, and M.~Vallisneri,
Phys.~Rev.~D {\bf 70}, 104003 (2004);\\
A.~Buonanno, Y.~Chen, Y.~Pan, H.~Tagoshi, and M.~Vallisneri,
Phys.~Rev.~D {\bf 72}, 084027 (2005) 

\bibitem{UsefulCycles}
T.~Damour, B.R.~Iyer, and B.S.~Sathyaprakash, 
Phys.~Rev.~D {\bf 62}, 084036 (2000)

\bibitem{SV} A.M.~Sintes and A.~Vecchio,
in \emph{Proceedings of the Rencontres de Moriond: Gravitational waves and experimental gravity}, 
ed.~J.~Dumarchez (Editions Fronti\`eres, Paris, 2000);\\
A.M.~Sintes and A.~Vecchio,
in \emph{Gravitational Waves, AIP Conference Proceedings Vol.~523}, ed.~S.~Meshkov (AIP Press, New York, 2000) 

\bibitem{letter} C.~Van Den Broeck, Class.~Quantum Grav.~{\bf 23}, L51-L58 (2006)

\bibitem{HellingsMoore}
T.A.~Moore and R.W.~Hellings,
Phys.~Rev.~D {\bf 65}, 062001 (2002);\\
R.W.~Hellings and T.A.~Moore,
Class.~Quantum Grav.~{\bf 20}, S181-S192 (2003)

\bibitem{Punturo} M.~Punturo, private communication (2006)

\bibitem{paramest} C.~Van Den Broeck and A.S.~Sengupta,
in preparation (2006)

\bibitem{Arunetal} K.G.~Arun, B.R.~Iyer, B.S.~Sathyaprakash and P.A.~Sundararajan,
Phys.~Rev. D {\bf 71}, 084008 (2005); Erratum ibid.~{\bf 72}, 069903 (2005)

\bibitem{SPA} K.S.~Thorne,
in \emph{300 Years of Gravitation}, eds.~S.W.~Hawking and W.~Israel (Cambridge University Press, Cambridge, 
England, 1987), p.~330;\\
B.S.~Sathyaprakash and S.V.~Dhurandhar,
Phys.~Rev.~D {\bf 44}, 3819-3934 (1991)

\bibitem{PRD66} T.~Damour, B.~Iyer, and B.S.~Sathyaprakash,
Phys.~Rev.~D {\bf 66}, 027502 (2002)

\bibitem{SPAvsFFT} S.~Droz, D.J.~Knapp, E.~Poisson, and B.J.~Owen,
Phys.~Rev.~D {\bf 59} 124016 (1999)

\bibitem{LSO} J.P.A.~Clark and D.M.~Eardley, 
Astrophys.~J.~{\bf 215}, 311 (1977);\\
J.K.~Blackburn and S.~Detweiler, 
Phys.~Rev.~D {\bf 46}, 2318 (1992);\\
G.B.~Cook,
Phys.~Rev.~D {\bf 50}, 5025 (1994);\\
L.E.~Kidder, C.M.~Will, and A.G.~Wiseman,
Class.~Quantum Grav.~{\bf 9}, L127 (1992);\\
L.E.~Kidder, C.M.~Will, and A.G.~Wiseman,
Phys.~Rev.~D {\bf 47}, 3281 (1993);\\
N.~Wex and G.~Sch\"afer,
Class.~Quantum Grav.~{\bf 10}, 2729 (1993);\\
G.~Sch\"afer and N.~Wex, 
Phys.~Lett.~A {\bf 174}, 196 (1993);\\
G.~Sch\"afer and N.~Wex,
Phys.~Lett.~A {\bf 177}, 461(E) (1993);\\
T.~Damour, B.~Iyer, and B.S.~Sathyaprakash,
Phys.~Rev.~D {\bf 57}, 885 (1998);\\
T.W.~Baumgarte,
Phys.~Rev.~D {\bf 62}, 024018 (2000);\\
T.~Damour, P.~Jaranowski, and G.~Sch\"afer,
Phys.~Rev.~D {\bf 62}, 084011 (2000)

\bibitem{QuinlanShapiro} G.D.~Quinlan and S.L.~Shapiro, 
Astrophys.~J.~{\bf 343}, 725-749 (1989);\\
G.D.~Quinlan and S.L.~Shapiro,
Ibid.~{\bf 356}, 483-500 (1990)

\bibitem{TutukovYungelson} A.V.~Tutukov and L.R.~Yungelson,
Mon.~Not.~R.~Astron.~Soc.~{\bf 260}, 675-678 (1993)

\bibitem{Lipunovetal} V.M.~Lipunov, K.A.~Postnov, and M.E.~Prokhorov,
New Astron.~{\bf 2}, 43-52 (1997)

\bibitem{SigurdssonHernquist} S.~Sigurdsson and L.~Hernquist,
Nature {\bf 364}, 423-425 (1993)

\bibitem{FlanaganHughes} \'E.\'E.~Flanagan and S.A.~Hughes,
Phys.~Rev.~D.~{\bf 57}, 4535-4565 (1998)

\bibitem{Kalogera} V.~Kalogera, private communication (2006)

\bibitem{WMAP} C.L.~Bennett et al.,
Astrophys.~J.~S.~{\bf 148}, 1 (2003)

\bibitem{GurselTinto} Y.~G\"ursel and M.~Tinto,
Phys.~Rev.~D {\bf 40}, 3884-3938 (1989)

\bibitem{Schutz} B.F.~Schutz, 
in \emph{Gravitational Wave Data Analysis}, 
ed.~B.F.~Schutz (Kluwer Academic Publishers, Dordrecht, 1989)

\bibitem{Schutz1} B.F.~Schutz, 
Class.~Quantum Grav.~{\bf 6}, 1761-1780 (1989)

\bibitem{Phinney} E.S.~Phinney,
Astrophys.~J.~{\bf 380}, L17-L21 (1991)

\bibitem{Paturel} G.~Paturel, P.~Fouque, L.~Bottinelli, and L.~Gouguenheim,
Astron.~Astrophys.~Suppl.~{\bf 80}, 299-315 (1989)

\bibitem{Tully} R.B.~Tully,
\emph{Nearby Galaxy Catalog} (Cambridge University Press, Cambridge, England, 1988)

\bibitem{Nutzmanetal} P.~Nutzman, V.~Kalogera, L.S.~Finn, C.~Hendrickson, and K.~Belczynski,
Astrophys.~J.~{\bf 612}, 364-374 (2004)

\bibitem{Phinney2} E.S.~Phinney,
astro-ph/0108028

\bibitem{Rates} L.S.~Finn,
Phys.~Rev.~D {\bf 53}, 2878-2894 (1996);\\
L.S.~Finn,
in \emph{AIP Conference Proceedings 575} (Melville, New York, 2001);\\
V.~Kalogera, R.~Narayan, D.N.~Spergel, J.H.~Taylor, 
Astrophys.~J.~{\bf 556}, 340-356 (2001)

\bibitem{BBW} E.~Berti, A.~Buonanno, and C.M.~Will,
Phys.~Rev.~D {\bf 71}, 084025 (2005);\\
E.~Berti, A.~Buonanno, and C.M.~Will,
Class.~Quantum Grav.~{\bf 22}, S943-S954 (2003)

\bibitem{Testing} K.~Arun, B.R.~Iyer, M.S.S.~Qusailah, and B.S.~Sathyaprakash,
Class.~Quantum Grav.~{\bf 23}, L37-L43 (2006);\\
K.G.~Arun, B.R.~Iyer, M.S.S.~Qusailah, and B.S.~Sathyaprakash, 
gr-qc/0604067

\bibitem{Courtyetal} J.-M.~Courty, A.~Heidmann and M.~Pinard, 
Phys.~Rev.~Lett.~{\bf 90} 083601 (2003);\\
J.-M.~Courty, A.~Heidmann and M.~Pinard, 
Europhys.~Lett.~{\bf 63} 226-232 (2003)





\end{thebibliography}
\end{document}